\documentclass[trackchanges, twocolumn]{aastex701}

\usepackage[normalem]{ulem}

\begin{document}

\title{Investigating IceCube Neutrino Alerts with the HAWC Gamma-Ray Observatory.}

\author{R.~Alfaro}
\email{}
\affiliation{{Instituto de F\'{i}sica, Universidad Nacional Autónoma de México, Ciudad de Mexico, Mexico }}

\author{C.~Alvarez}
\email{}
\address{Universidad Autónoma de Chiapas, Tuxtla Gutiérrez, Chiapas, México}

\author{A.~Andrés}
\email{}
\address{Instituto de Astronom\'{i}a, Universidad Nacional Autónoma de México, Ciudad de Mexico, Mexico }

\author{E.~Anita-Rangel}
\email{}
\address{Instituto de Astronom\'{i}a, Universidad Nacional Autónoma de México, Ciudad de Mexico, Mexico }

\author[0000-0002-0595-9267]{M.~Araya}
\email{}
\address{Universidad de Costa Rica, San José 2060, Costa Rica}

\author[0009-0005-5590-8563]{J.C.~Arteaga-Velázquez}
\email{}
\address{Universidad Michoacana de San Nicolás de Hidalgo, Morelia, Mexico }

\author[0000-0002-4020-4142]{D.~Avila Rojas}
\email{}
\address{Instituto de Astronom\'{i}a, Universidad Nacional Autónoma de México, Ciudad de Mexico, Mexico }

\author[0000-0002-2084-5049]{H.A.~Ayala Solares}
\email{}
\address{Temple University, Department of Physics, 1925 N. 12th Street, Philadelphia, PA 19122, USA}

\author[0000-0002-5529-6780]{R.~Babu}
\email{}
\address{Department of Physics and Astronomy, Michigan State University, East Lansing, MI, USA }

\author[0000-0003-3207-105X]{E.~Belmont-Moreno}
\email{}
\affiliation{{Instituto de F\'{i}sica, Universidad Nacional Autónoma de México, Ciudad de Mexico, Mexico }}

\author{A.~Bernal}
\email{}
\address{Instituto de Astronom\'{i}a, Universidad Nacional Autónoma de México, Ciudad de Mexico, Mexico }

\author[0000-0002-4042-3855]{K.S.~Caballero-Mora}
\email{}
\address{Universidad Autónoma de Chiapas, Tuxtla Gutiérrez, Chiapas, México}

\author[0000-0003-2158-2292]{T.~Capistrán}
\email{}
\address{Instituto Nacional de Astrof\'{i}sica, Óptica y Electrónica, Puebla, Mexico }

\author[0000-0002-8553-3302]{A. Carramiñana}
\email{alberto@inaoep.mx}
\address{Instituto Nacional de Astrof\'{i}sica, Óptica y Electrónica, Puebla, Mexico }

\author{F.~Carreón}
\email{}
\address{Instituto de Astronom\'{i}a, Universidad Nacional Autónoma de México, Ciudad de Mexico, Mexico }

\author[0000-0002-6144-9122]{S.~Casanova}
\email{}
\address{Institute of Nuclear Physics Polish Academy of Sciences, PL-31342 IFJ-PAN, Krakow, Poland }

\author[0000-0002-1132-871X]{J.~Cotzomi}
\email{}
\address{Facultad de Ciencias F\'{i}sico Matemáticas, Benemérita Universidad Autónoma de Puebla, Puebla, Mexico }

\author[0000-0002-7747-754X]{S.~Coutiño de León}
\email{}
\address{Instituto de Física Corpuscular, CSIC, Universitat de València, E-46980, Paterna, Valencia, Spain}

\author[0000-0002-8528-9573]{C.~de León}
\email{}
\address{Instituto de Física Corpuscular, CSIC, Universitat de València, E-46980, Paterna, Valencia, Spain }

\author[0000-0001-9643-4134]{E.~De la Fuente}
\email{}
\address{Departamento de F\'{i}sica, Centro Universitario de Ciencias Exactase Ingenierias, Universidad de Guadalajara, Guadalajara, Mexico }

\author{P.~Desiati}
\email{}
\address{Dept. of Physics and Wisconsin IceCube Particle Astrophysics Center, University of Wisconsin{\textemdash}Madison, Madison, WI, USA}

\author[0000-0002-7574-1298]{N.~Di Lalla}
\email{}
\address{Department of Physics, Stanford University: Stanford, CA 94305-4060, USA}

\author{R.~Diaz Hernandez}
\email{}
\address{Instituto Nacional de Astrof\'{i}sica, Óptica y Electrónica, Puebla, Mexico }

\author[0000-0002-2987-9691]{M.A.~DuVernois}
\email{}
\address{Dept. of Physics and Wisconsin IceCube Particle Astrophysics Center, University of Wisconsin{\textemdash}Madison, Madison, WI, USA}

\author[0000-0002-0087-0693]{J.C.~Díaz-Vélez}
\email{}
\address{Dept. of Physics and Wisconsin IceCube Particle Astrophysics Center, University of Wisconsin{\textemdash}Madison, Madison, WI, USA}

\author[0000-0001-5737-1820]{K.~Engel}
\email{}
\address{Department of Physics, University of Maryland, College Park, MD, USA }

\author[0000-0001-7074-1726]{C.~Espinoza}
\email{}
\affiliation{{Instituto de F\'{i}sica, Universidad Nacional Autónoma de México, Ciudad de Mexico, Mexico }}

\author[0000-0002-0173-6453]{N.~Fraija}
\email{}
\address{Instituto de Astronom\'{i}a, Universidad Nacional Autónoma de México, Ciudad de Mexico, Mexico }

\author{S.~Fraija}
\email{}
\address{Instituto de Astronom\'{i}a, Universidad Nacional Autónoma de México, Ciudad de Mexico, Mexico }

\author{A.~Galván-Gámez}
\email{}
\affiliation{{Instituto de F\'{i}sica, Universidad Nacional Autónoma de México, Ciudad de Mexico, Mexico }}

\author[0000-0002-4188-5584]{J.A.~García-González}
\email{}
\address{Tecnologico de Monterrey, Escuela de Ingenier\'{i}a y Ciencias, Ave. Eugenio Garza Sada 2501, Monterrey, N.L., Mexico, 64849}

\author[0000-0003-1122-4168]{F.~Garfias}
\email{}
\address{Instituto de Astronom\'{i}a, Universidad Nacional Autónoma de México, Ciudad de Mexico, Mexico }

\author{N.~Ghosh}
\email{}
\address{Department of Physics, Michigan Technological University, Houghton, MI, USA }

\author{A.~Gonzalez Muñoz}
\email{}
\affiliation{{Instituto de F\'{i}sica, Universidad Nacional Autónoma de México, Ciudad de Mexico, Mexico }}

\author{M.M.~González}
\email{}
\address{Instituto de Astronom\'{i}a, Universidad Nacional Autónoma de México, Ciudad de Mexico, Mexico }

\author{J.A.~González}
\email{}
\address{Universidad Michoacana de San Nicolás de Hidalgo, Morelia, Mexico }

\author[0000-0002-9790-1299]{J.A.~Goodman}
\email{}
\address{Department of Physics, University of Maryland, College Park, MD, USA }

\author[0000-0002-0870-2328]{D.~Guevel}
\email{}
\address{Department of Physics, Michigan Technological University, Houghton, MI, USA }

\author{J.~Gyeong}
\email{}
\address{Department of Physics, Sungkyunkwan University, Suwon 16419, South Korea}

\author[0000-0001-9844-2648]{J.P.~Harding}
\email{}
\address{Los Alamos National Laboratory, Los Alamos, NM, USA }

\author{S.~Hernández-Cadena}
\email{}
\address{Tsung-Dao Lee Institute \& School of Physics and Astronomy, Shanghai Jiao Tong University, 800 Dongchuan Rd, Shanghai, SH 200240, China}

\correspondingauthor {Ian Herzog}
\author[0000-0001-5169-723X]{I. Herzog} 
\email[show]{herzogia@msu.edu}
\address{Department of Physics and Astronomy, Michigan State University, East Lansing, MI, USA }

\author[0000-0002-1031-7760]{J.~Hinton}
\email{}
\address{Max-Planck Institute for Nuclear Physics, 69117 Heidelberg, Germany}

\author[0000-0002-5447-1786]{D.~Huang}
\email{}
\address{Department of Physics and Astronomy, University of Delaware, Newark, DE, USA}

\author[0000-0002-5527-7141]{F.~Hueyotl-Zahuantitla}
\email{}
\address{Universidad Autónoma de Chiapas, Tuxtla Gutiérrez, Chiapas, México}

\author[0000-0002-3302-7897]{P.~Hüntemeyer}
\email{}
\address{Department of Physics, Michigan Technological University, Houghton, MI, USA }

\author[0000-0001-5811-5167]{A.~Iriarte}
\email{}
\address{Instituto de Astronom\'{i}a, Universidad Nacional Autónoma de México, Ciudad de Mexico, Mexico }

\author{S.~Kaufmann}
\email{}
\address{Universidad Politecnica de Pachuca, Pachuca, Hgo, Mexico }

\author[0000-0003-4785-0101]{D.~Kieda}
\email{}
\address{Department of Physics and Astronomy, University of Utah, Salt Lake City, UT, USA }

\author{K.~Leavitt}
\email{}
\address{Department of Physics, Michigan Technological University, Houghton, MI, USA }

\author[0000-0002-2153-1519]{J.~Lee}
\email{}
\address{University of Seoul, Seoul, Rep. of Korea}

\author[0000-0002-2467-5673]{W.H.~Lee}
\email{}
\address{Instituto de Astronom\'{i}a, Universidad Nacional Autónoma de México, Ciudad de Mexico, Mexico }

\author[0000-0001-5516-4975]{H.~León Vargas}
\email{}
\affiliation{{Instituto de F\'{i}sica, Universidad Nacional Autónoma de México, Ciudad de Mexico, Mexico }}

\author[0000-0003-2696-947X]{J.T.~Linnemann}
\email{}
\address{Department of Physics and Astronomy, Michigan State University, East Lansing, MI, USA }

\author[0000-0001-8825-3624]{A.L.~Longinotti}
\email{}
\address{Instituto de Astronom\'{i}a, Universidad Nacional Autónoma de México, Ciudad de Mexico, Mexico }

\author[0000-0003-2810-4867]{G.~Luis-Raya}
\email{}
\address{Universidad Politecnica de Pachuca, Pachuca, Hgo, Mexico }

\author[0000-0001-8088-400X]{K.~Malone}
\email{}
\address{Los Alamos National Laboratory, Los Alamos, NM, USA }

\author[0000-0001-9052-856X]{O.~Martinez}
\email{}
\address{Facultad de Ciencias F\'{i}sico Matemáticas, Benemérita Universidad Autónoma de Puebla, Puebla, Mexico }

\author[0000-0002-2824-3544]{J.~Martínez-Castro}
\email{}
\address{Centro de Investigaci\'on en Computaci\'on, Instituto Polit\'ecnico Nacional, M\'exico City, M\'exico.}

\author[0000-0002-2610-863X]{J.A.~Matthews}
\email{}
\address{Dept of Physics and Astronomy, University of New Mexico, Albuquerque, NM, USA }

\author[0000-0002-8390-9011]{P.~Miranda-Romagnoli}
\email{}
\address{Universidad Autónoma del Estado de Hidalgo, Pachuca, Mexico }

\author{P.E.~Mirón-Enriquez}
\email{}
\address{Instituto de Astronom\'{i}a, Universidad Nacional Autónoma de México, Ciudad de Mexico, Mexico }

\author[0000-0003-3207-105X]{E.~Moreno}
\email{}
\address{Facultad de Ciencias F\'{i}sico Matemáticas, Benemérita Universidad Autónoma de Puebla, Puebla, Mexico }

\author[0000-0002-7675-4656]{M.~Mostafá}
\email{}
\address{Temple University, Department of Physics, 1925 N. 12th Street, Philadelphia, PA 19122, USA}

\author{M.~Najafi}
\email{}
\address{Department of Physics, Michigan Technological University, Houghton, MI, USA }

\author[0000-0003-0587-4324]{A.~Nayerhoda}
\email{}
\address{Institute of Nuclear Physics Polish Academy of Sciences, PL-31342 IFJ-PAN, Krakow, Poland }

\author[0000-0003-1059-8731]{L.~Nellen}
\email{}
\address{Instituto de Ciencias Nucleares, Universidad Nacional Autónoma de Mexico, Ciudad de Mexico, Mexico }

\author[0000-0002-6859-3944]{M.U.~Nisa}
\email{}
\address{Department of Physics and Astronomy, Michigan State University, East Lansing, MI, USA }

\author[0000-0001-7099-108X]{R.~Noriega-Papaqui}
\email{}
\address{Universidad Autónoma del Estado de Hidalgo, Pachuca, Mexico }

\author[0000-0002-5448-7577]{N.~Omodei}
\email{}
\address{Department of Physics, Stanford University: Stanford, CA 94305-4060, USA}

\author{M.~Osorio-Archila}
\email{}
\address{Instituto de Astronom\'{i}a, Universidad Nacional Autónoma de México, Ciudad de Mexico, Mexico }

\author{E.~Ponce}
\email{}
\address{Facultad de Ciencias F\'{i}sico Matemáticas, Benemérita Universidad Autónoma de Puebla, Puebla, Mexico }

\author[0000-0002-8774-8147]{Y.~Pérez Araujo}
\email{}
\affiliation{{Instituto de F\'{i}sica, Universidad Nacional Autónoma de México, Ciudad de Mexico, Mexico }}

\author[0000-0001-5998-4938]{E.G.~Pérez-Pérez}
\email{}
\address{Universidad Politecnica de Pachuca, Pachuca, Hgo, Mexico }

\author[0000-0002-6524-9769]{C.D.~Rho}
\email{}
\address{Department of Physics, Sungkyunkwan University, Suwon 16419, South Korea}

\author[0000-0003-1327-0838]{D.~Rosa-González}
\email{}
\address{Instituto Nacional de Astrof\'{i}sica, Óptica y Electrónica, Puebla, Mexico }

\author[0000-0002-4204-5026]{M.~Roth}
\email{}
\address{Los Alamos National Laboratory, Los Alamos, NM, USA }

\author{H.~Salazar}
\email{}
\address{Facultad de Ciencias F\'{i}sico Matemáticas, Benemérita Universidad Autónoma de Puebla, Puebla, Mexico }

\author{D.~Salazar-Gallegos}
\email{}
\address{Department of Physics and Astronomy, Michigan State University, East Lansing, MI, USA }

\author[0000-0001-6079-2722]{A.~Sandoval}
\email{}
\affiliation{{Instituto de F\'{i}sica, Universidad Nacional Autónoma de México, Ciudad de Mexico, Mexico }}

\author[0000-0001-8644-4734]{M.~Schneider}
\email{}
\address{Department of Physics, University of Maryland, College Park, MD, USA }

\author{J.~Serna-Franco}
\email{}
\affiliation{{Instituto de F\'{i}sica, Universidad Nacional Autónoma de México, Ciudad de Mexico, Mexico }}

\author{M.~Shin}
\email{}
\address{Department of Physics, Sungkyunkwan University, Suwon 16419, South Korea}

\author[0000-0002-1012-0431]{A.J.~Smith}
\email{}
\address{Department of Physics, University of Maryland, College Park, MD, USA }

\author{Y.~Son}
\email{}
\address{University of Seoul, Seoul, Rep. of Korea}

\author[0000-0002-1492-0380]{R.W.~Springer}
\email{}
\address{Department of Physics and Astronomy, University of Utah, Salt Lake City, UT, USA }

\author{O.~Tibolla}
\email{}
\address{Universidad Politecnica de Pachuca, Pachuca, Hgo, Mexico }

\author[0000-0001-9725-1479]{K.~Tollefson}
\email{}
\address{Department of Physics and Astronomy, Michigan State University, East Lansing, MI, USA }

\author[0000-0002-7102-3352]{I.~Torres}
\email{}
\address{Instituto Nacional de Astrof\'{i}sica, Óptica y Electrónica, Puebla, Mexico }

\author[0000-0002-7102-3352]{R.~Torres-Escobedo}
\email{}
\address{Tsung-Dao Lee Institute \& School of Physics and Astronomy, Shanghai Jiao Tong University, 800 Dongchuan Rd, Shanghai, SH 200240, China}

\author[0000-0003-0715-7513]{E.~Varela}
\email{}
\address{Facultad de Ciencias F\'{i}sico Matemáticas, Benemérita Universidad Autónoma de Puebla, Puebla, Mexico }

\author[0000-0001-6876-2800]{L.~Villaseñor}
\email{}
\address{Facultad de Ciencias F\'{i}sico Matemáticas, Benemérita Universidad Autónoma de Puebla, Puebla, Mexico }

\author[0000-0001-6798-353X]{X.~Wang}
\email{}
\address{Department of Physics, Missouri University of Science and Technology, Rolla, MO, US}

\author{Z.~Wang}
\email{}
\address{Department of Physics, Missouri University of Science and Technology, Rolla, MO, US}

\author[0000-0003-2141-3413]{I.J.~Watson}
\email{}
\address{University of Seoul, Seoul, Rep. of Korea}

\author[0009-0005-7243-1402]{H.~Wu}
\email{}
\address{Dept. of Physics and Wisconsin IceCube Particle Astrophysics Center, University of Wisconsin{\textemdash}Madison, Madison, WI, USA}

\author[0009-0006-3520-3993]{S.~Yu}
\email{}
\address{Department of Physics, Pennsylvania State University, University Park, PA, USA }

\author[0000-0003-0513-3841]{H.~Zhou}
\email{}
\address{Tsung-Dao Lee Institute \& School of Physics and Astronomy, Shanghai Jiao Tong University, 800 Dongchuan Rd, Shanghai, SH 200240, China}

\collaboration{all}{The HAWC collaboration}

\begin{abstract}

    Neutrino emission from astrophysical sources has long been considered a signature of cosmic-ray acceleration. The IceCube Neutrino Observatory has observed a diffuse flux of TeV-PeV neutrinos, but very few confirmed sources have emerged. With the recent publication of the IceCube Event Catalog (IceCat-1), IceCube has released a list of probable astrophysical neutrinos from May 2011 to December 2020, with follow-up alerts released via the Astrophysical Multi-messenger Observatory Network (AMON). Using the archival data from the High Altitude Water Cherenkov (HAWC) Gamma-Ray Observatory, we search for gamma-ray outbursts spatially coincident with these neutrino alerts using a Bayesian Block algorithm. In this work, we consider 380 alerts, up to May 19, 2026, that are within HAWC's field of view.  We observe approximately a 5\% coincident detection rate, which is consistent with expectations from background.  Two of these detections contain the Active Galactic Nuclei (AGN) Markarian 421 and Markarian 501. We discuss the possibility of these AGN producing the observed neutrino events when considering the HAWC results.

\end{abstract}

\section{Introduction}\label{sec:intro}

Multi-messenger astronomy has become essential for connecting observations across the electromagnetic (EM) spectrum with signals from gravitational waves and particles like neutrinos. Specifically, studies in the TeV range can help place limits on particle energies and offer clues about the underlying acceleration mechanisms of ultra high-energy cosmic rays.  Multi-wavelength observations of events like the joint EM and gravitational wave observations of GW150914 \citep{GW_example} and the joint neutrino-EM flaring of TXS 0506+056 \citep{TEXAS_og},have significantly improved our understanding of the physics of extremely energetic environments.

In particular, coincident neutrino and gamma-ray emission can be used to place constraints or determine the origin of the most energetic cosmic rays.  In the TeV regime, gamma-rays are produced one of two ways: either through the inverse Compton scattering of leptons or pion decay in hadronic interactions \citep{aharonianbook}.  Crucially, pion decay produces neutrinos in addition to gamma-rays.  Therefore, observing joint neutrino and TeV gamma-ray emission from sources would provide conclusive evidence about the nature of a particular gamma-ray source.

There have been few neutrino sources identified as of this writing.  The IceCube Neutrino Observatory (further discussed in Section \ref{sec:icecube}), is sensitive to TeV-PeV neutrinos and has recently observed diffuse neutrino emission from the galactic plane \citep{IC_galactic} along with a few extragalactic sources.  These include Active Galactic Nuclei (AGN) like TXS 0506+056 \citep{TEXAS_og} and NGC 1068 \citep{ngc_1068}. 

Additionally, IceCube sends out public alerts for probable astrophysical neutrinos.  Archival versions of these alerts were published as IceCat-1 \citep{icat1} and current alerts are distributed via the Astrophysical Multi-messenger Observatory Network (AMON) \citep{amon} and NASA's General Coordinate Network (GCN).  These can be found on AMON's website \footnote{\url{https://gcn.gsfc.nasa.gov/amon_icecube_gold_bronze_events.html}} (see Section \ref{sec:datasets}).

As for TeV gamma-ray emission, current wide-field observatories like the High Altitude Water Cherenkov (HAWC) Gamma-ray Observatory \citep{hawc_nim}, the Large High Altitude Air Shower Observatory (LHAASO) \citep{lhaaso}, and future observatories like the Southern Wide-field Gamma-ray Observatory (SWGO) \citep{swgo} are ideally suited for observing both steady-state and transient sources.  HAWC has observed the two flaring AGN Markarians (Mrks) 421 and 501 \citep{hawc_mrk_steadystate} along with several other AGN \citep{hawcradio}.  Additionally, HAWC is sensitive to transient events \citep{hawc_monitoring} and can be used to follow up any alerts sent by IceCube.

There are many possible models that predict joint gamma-ray and neutrino emission simultaneously. Many such models are explored in \cite{texas_multi}.  These include both lepto-hadronic and hadronic models.  The former considers electrons producing both the synchrotron and high-energy peaks along with neutrinos while suppressing TeV hadronic photons; meanwhile, the latter still considers electrons producing the synchrotron bump but hadronic interactions in the photomeson process produce the high-energy photons and neutrinos. Additionally, there are models that predict either delayed photon \citep{gamma_delayed_emission} or neutrino \citep{neutrino_delayed_emission} emission. These are highly dependent on the unique environments each of these events occur within.  All of these models require extensive multi-wavelength data to properly constrain and are broadly beyond the scope of this work. Section \ref{sec:multipain} explores a basic neutrino emission model based of HAWC observations of Mrks 421 and 501.

In this paper, we search for coincident flaring of gamma-rays with IceCube alerts in archival HAWC data. First, we introduce both the HAWC gamma-ray and IceCube neutrino observatories (Section \ref{sec:observatories}) and describe the data that are used from both of them in this analysis (Section \ref{sec:datasets}). In Section \ref{sec:method}, we discuss HAWC's data collection method and introduce the fitting algorithm: the Bayesian Block algorithm.  The results of applying this algorithm to HAWC's data are presented in Section \ref{sec:results}.  Next, a multi-messenger interpretation of the observed neutrino/gamma-ray coincidences is presented in Section \ref{sec:multipain} to test for possible location-coincident neutrino emission from AGN.  Systematic effects are then discussed in Section \ref{sec:discussion}.  

\section{Observatories}\label{sec:observatories}

\subsection{HAWC}

HAWC is located on the slopes of the Sierra Negra volcano in Mexico, at an altitude of about 4,100 meters. It is composed of 300 large water tanks, each equipped with photomultiplier tubes (PMTs) at the bottom to detect Cherenkov light generated by secondary particles from gamma-ray air showers. The main array contains 1,200 PMTs spread across the tanks, enabling a wide Field of View (FOV). HAWC observes gamma rays in the energy range from about 300 GeV up to beyond 100 TeV.  With an uptime of 93\% and 2 sr FOV, it is ideal for capturing high-energy transient gamma-ray events \citep{hawc-nim-paper}.

\subsection{IceCube}\label{sec:icecube}

The IceCube Neutrino Observatory, located at the South Pole, is a cubic-kilometer-scale detector embedded deep within the Antarctic ice. It consists of three main components: the main in-ice array, a cosmic ray detector called IceTop, and a denser sub-array called DeepCore. The in-ice array is made up of 86 vertical strings, with each containing Digital Optical Modules (DOMs), totaling 5,160 DOMs.  Each DOM contains a PMT that detects the Cherenkov light produced when neutrinos interact with the ice. IceCube is sensitive to a broad energy range, from roughly 10 GeV with DeepCore up to beyond 1 PeV. The detector has an uptime of $>99\%$ and a 360 degree FOV, making it ideal for capturing transient neutrino events \citep{icecube-hardware}. 

\section{Data}\label{sec:datasets}

The IceCube neutrino data considered are astrophysical neutrino alerts from \cite{icat1} from 2011 to 2020 and the public alerts distributed via GCN onwards.  This analysis considers alerts up to May 19, 2026, with 380 alerts in HAWC's FOV.  The alerts contain the alert name, time, location (including 95\% containment), energy, and probability of being astrophysical.  They are split into two categories: bronze if the probability is $\ge 30\%$ and gold if the probability is $\ge50\%$.  These are indicated by either B or G, respectively. Lastly, IceCube's Point Spread Function (PSF) quite variable depending on the alert, ranging from $>$10 to $<$0.2 degrees.

\begin{figure*}[ht!]
    \centering
    \includegraphics[width=1.0\linewidth]{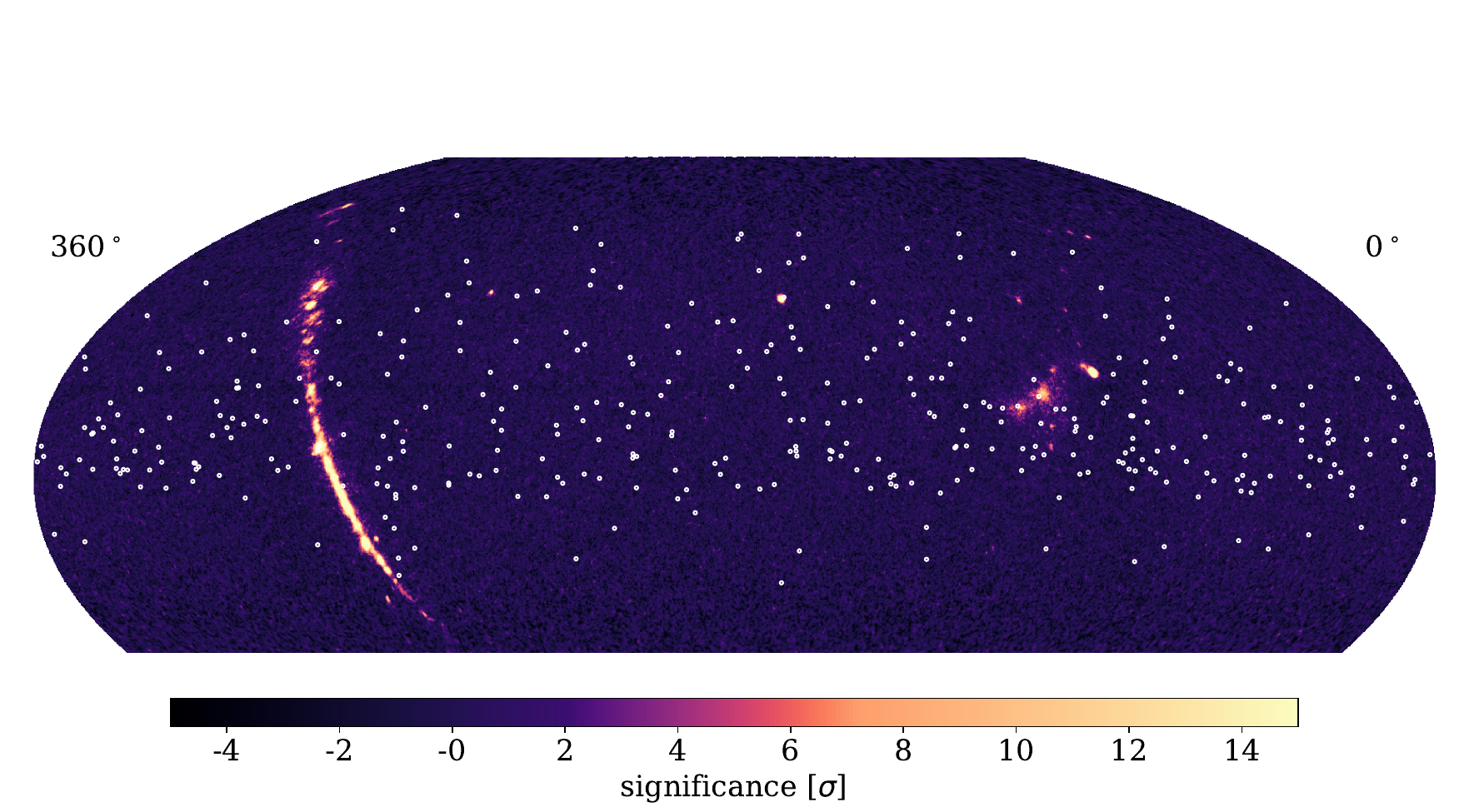}
    \caption{HAWC's full sky data map with the 380 visible IceCube alerts.  The white dots show the alerts that lie within the declination interval $-20^\circ$ to $60^\circ$.  This range is selected to constrain large IceCube alerts and prevent the ROI from extending beyond HAWC's FOV.  The plot is drawn in celestial coordinates with an index of 2.7 and a pivot energy ($E_p$) of 7 TeV.}
    \label{fig:ic_alert_dist}
\end{figure*}

For HAWC data, observations between March 20, 2015 and January 15, 2024 are considered.  These are grouped into two data sets: a time-integrated map and an ensemble of daily maps.  The former contains all days in the observation interval that have both a full transit and no known significant detector effects (maintenance, strong electric fields, etc) as these can affect data quality.  The latter contains every day of data HAWC collected, even if some data is missing or is affected by aforementioned detector effects.  As such, the time-integrated set contains 2565 days of data \citep{Crab_2024} and the daily map set contains 3175 days of data, though this number will vary depending on the selected IceCube alert. Locational coincidences are searched for within this range, with potential temporal correlations also considered.  The implications of potential detector effects in the daily maps are discussed in Section \ref{sec:discussion}.  

The full data set is used to search for potential neutrino coincidences with HAWC steady-state sources while the daily maps can search for transient events like potential flares.  All the IceCube alerts visible to HAWC are shown in Figure \ref{fig:ic_alert_dist}. As with IceCube, HAWC's PSF is dependent on the events observed and can range from $>$1 to $\approx$0.1 degrees for lower- and higher-energy events, respectively.

HAWC utilizes three different techniques to reconstruct its data. Discussed in \cite{Crab_2024}, these are the $f_{hit}$, Neural Network (NN), and Ground Parameter (GP) estimators.  In brief, the $f_{hit}$ dataset is constructed by recording the fraction of PMTs triggered by an air shower and is good for low- to medium-energy events (0.3 to $\approx 10$ TeV). Above 10 TeV, the detector becomes saturated and triggers all PMTs, making accurate energy estimation difficult.  The NN and GP are both energy estimators that span HAWC's full energy range (0.3 to $>100$ TeV).  The NN uses primarily the fractional charge distribution of nine concentric annuli around a shower core to reconstruct the principle energy.  The GP estimator fits the lateral distribution of the showers to reconstruct the initial gamma ray.  HAWC's daily maps utilize the $f_{hit}$ estimator while the reconstructed data set makes three separate maps, one for each estimator. The $f_{hit}$ and NN maps are presented in this work, while the GP was used to validation.

When creating HAWC's significance maps, a likelihood test is performed by defining a test statistic (TS) that is the ratio of an alternate hypothesis (background + source) and null hypothesis (background only).  The TS is defined as 

\begin{equation}
    \text{TS} = 2\ln \frac{L_{alt}}{L_{null}}.
\end{equation}

\noindent Assuming Wilks' theorem \citep{wilks1938large}, if the difference in free parameters between the two models is one, then the TS is distributed as a $\chi^2$ distribution with the degrees of freedom equal to the number of free parameters. The significance in deviation between the two models can be approximated as $\sigma = \sqrt{\text{TS}}$. 

The spectral assumption used for the model discussed above is a simple power law, given by 

\begin{equation}
    \frac{dE}{dN} = N_o \left( \frac{E}{E_p}\right)^{-\alpha},
\end{equation}

\noindent where $N_o$ is the flux normalization, $\alpha$ is the spectral index, and $E_p=7$ TeV is the pivot energy.  We created two maps with fixed alphas 2.0 and 3.0, allowing only the flux normalization to float.  These favor either harder indices/higher energy or softer indices/lower-energy events, respectively.  Henceforth, these maps will be referred to as 2.0 or 3.0 maps. 

\section{Methodology}\label{sec:method}

\subsection{Data Collection}\label{sec:data_collect}

Searching the HAWC data using the IceCube alerts is done as follows: first, the location and uncertainty region from each alert is extracted and used to construct a Region of Interest (ROI).  Then, the time integrated HAWC data set is scanned to find the brightest significance spot inside the ROI to determine if a steady gamma-ray source is present.  These results are used to draw potential correlations between the alerts and known HAWC sources (\ref{sec:multipain}).

\begin{figure*}[ht!]
    \centering
    \includegraphics[width=0.49\linewidth]{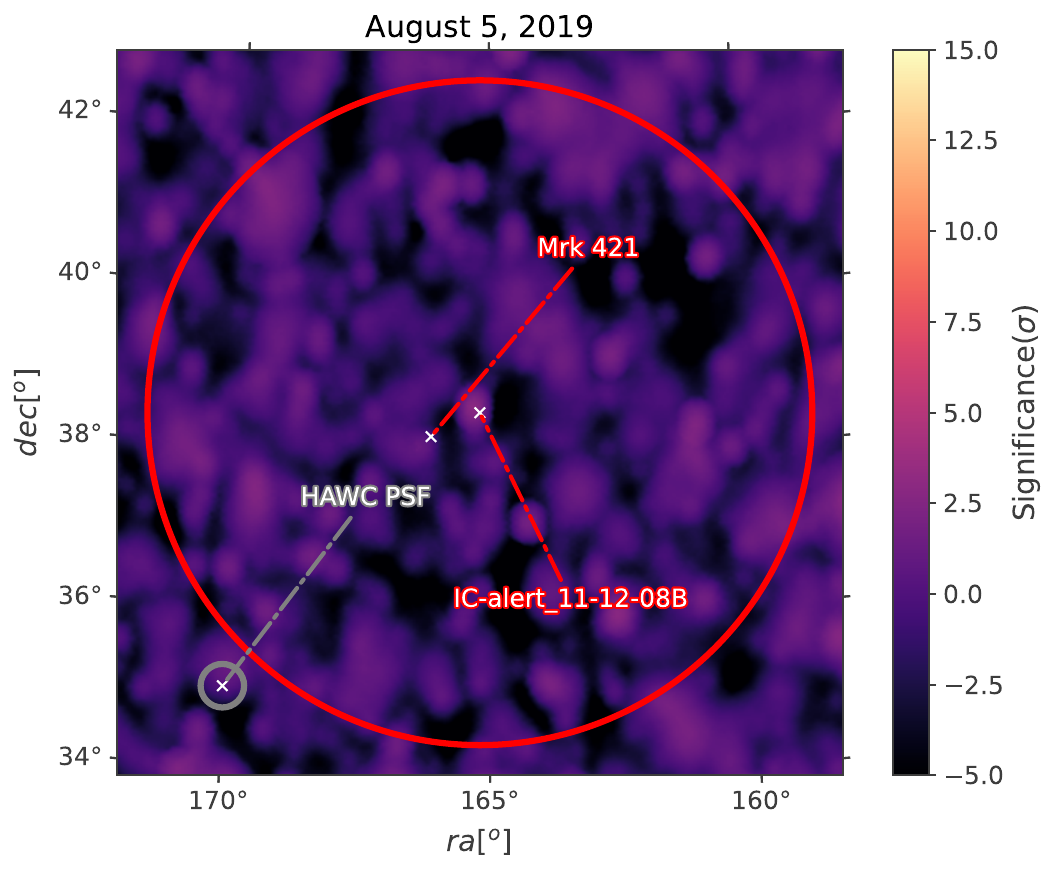}\hfill
    \includegraphics[width=0.49\linewidth]{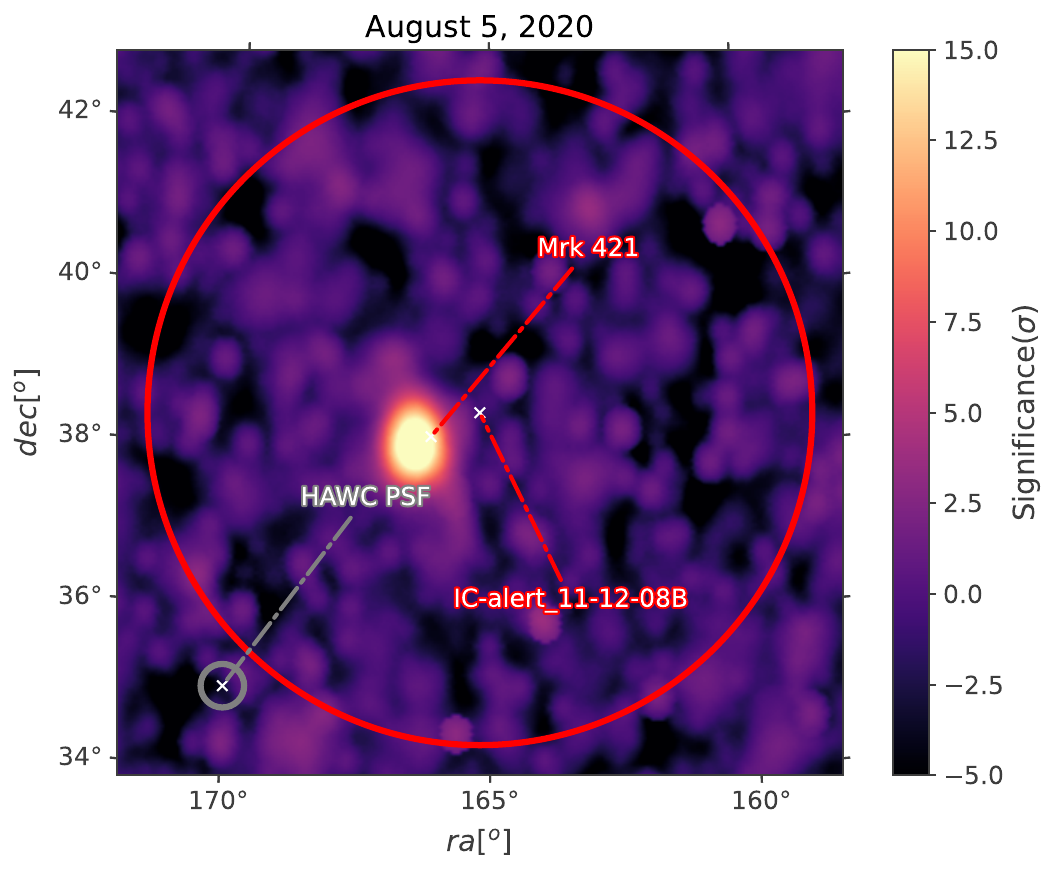}\hfill
    \includegraphics[width=0.65\linewidth]{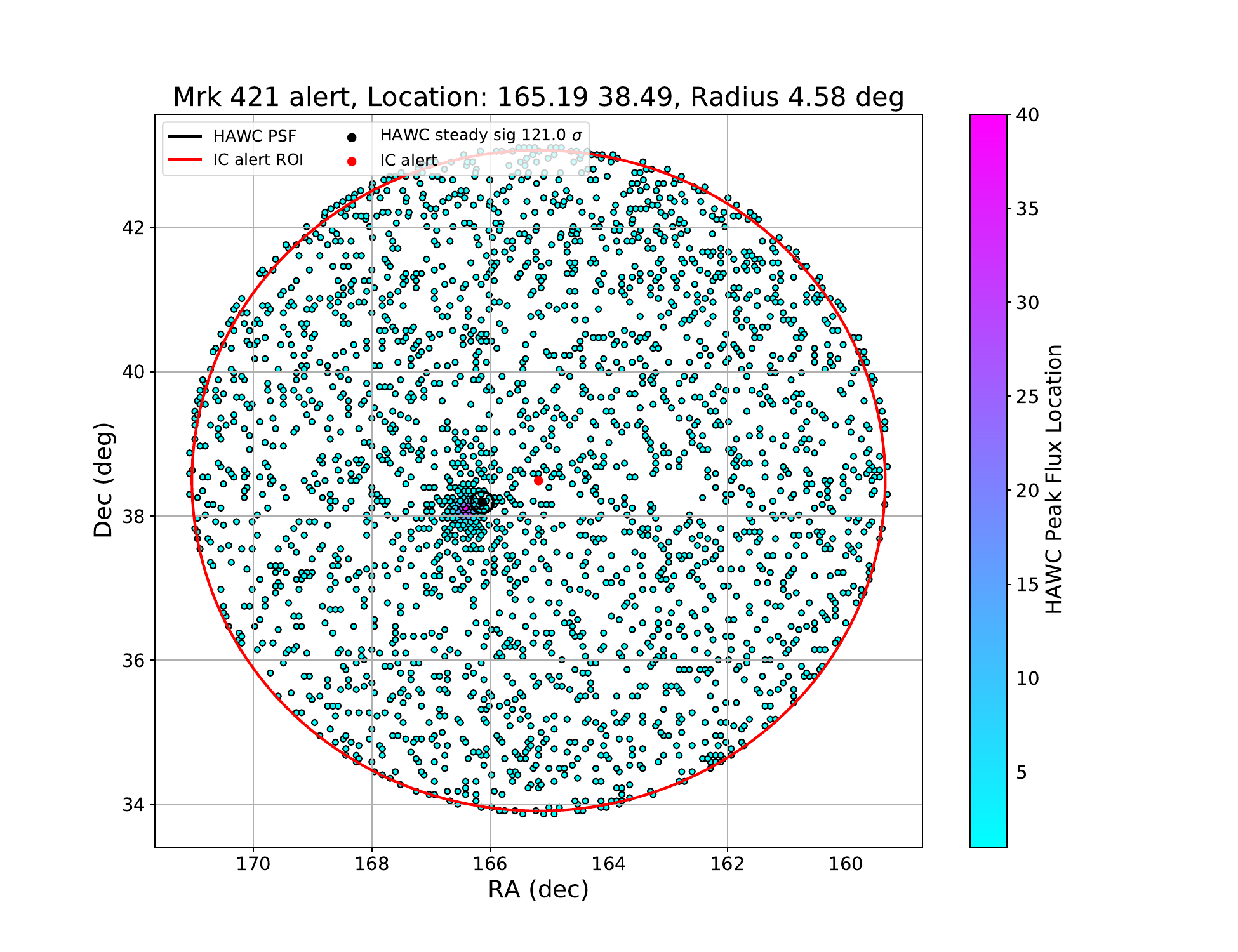}\hfill
    \caption{The top two plots indicate two daily maps of Mrk 421, one where it is quiet (LEFT) and when it is flaring (RIGHT).  The red circle indicates the IceCube alert's ROI.  HAWC's PSF is indicated by the grey circle in the lower left.  The bottom plot shows the frequency that a given RA, Dec has the max flux in the daily map. HAWC sees Mrk 421 to $121\sigma$ with the time-integrated data set.}
    \label{fig:mrkdata}
\end{figure*}

Next, both the 2.0 and 3.0 daily map sets are considered.  For each day, the pixels in the data maps within the ROI are scanned to search for the pixel with the maximum flux value, along with its corresponding uncertainty and location (RA, DEC).  If a flare (event above background) is present and visible to HAWC, its flux would appear above the background for the given day.  These four parameters are extracted, and then a light curve is made for the ROI. The TeV-bright AGN Mrk 421 is located 0.8 degrees from the neutrino alert IC111208B and is used to show this procedure.  Example significance maps and resulting maximum flux distribution are shown in Figure \ref{fig:mrkdata}.

\subsection{Bayesian Block Algorithm (BBA)}\label{sec:bba}

While extraordinarily bright transient sources like Mrk 421 can be found easily by eye, most sources are too dim to do this.  Therefore, we apply the Bayesian Block Algorithm (BBA), mathematically described in \cite{BB_paper}, and use the code implementation from \cite{BB_code}.  A brief description of the BBA is presented here.  The goal of the BBA is to find statistically significant deviations in data.  We use it in its ``piecewise function'' model with the algorithm set to the ``Point Measurements'' setting (see Section 3.3 in \cite{BB_paper}). The blocks are described by step functions with 0 slope spanning a certain interval determined by the sensitivity of the algorithm. The data assumes time-independent (1 value per time step) observations with some uncertainty.  

Each block is defined by ``change points'' at its start and end.  For a given dataset, the minimum number of change points is 2, with the dataset being fit with a single block.  A detection is defined by any fit that contains more than 2 change points. The parameter $n_{cp}$ gives the number of change points above background.  Additionally, the location and data values that resulted in the various change points can be saved. These are used later in Section \ref{sec:results}.

To determine the total number of blocks for a given data set, a global ``fitness function'' is defined as 

\begin{equation}
    F[P(T)] = \sum_{k=1}^{N_{blocks}} f(B_k),
\end{equation}

\noindent where the global fitness function $F$ for a given partition of the data $P(T)$ is the sum of each block's fitness functions $B_k$.  If we assume a steady-state emission model \mbox{$\lambda=s$}, with $\lambda$ being the model and $s$ some constant value, and a normal distribution with zero mean and a known variance for the observational error, the individual block fitness functions become a log likelihood function of form

\begin{equation}
    \log L_{\text{max}}^k = \sum_{n} \frac{x_n^2}{\sigma_n^2}.
\end{equation}

\noindent The block's fitness function is only dependent on what data and uncertainty $x_n \pm \sigma_n$ is included inside it and is time-independent.  To prevent overfitting, a penalty factor is included to penalize models with more blocks.  The new global fitness function now becomes

\begin{equation}
    F[P(T)] = \sum_{k=1}^{N_{blocks}} \left(\sum_{n} \frac{x_n^2}{\sigma_n^2}\right) - N_{blocks}\log \gamma,
\end{equation}

\noindent with $-\log \gamma$, also called the \emph{ncp\_prior}, being the penalty factor and is a hyperparameter.  The \emph{ncp\_prior} determines how sensitive the BBA is to deviations in the data.  If it is too small, the BBA will fit blocks to random noise fluctuations in the data while, if it is too large, it will not fit potential signals or flares.  Therefore, it must be calibrated before the BBA can be used in this analysis.

\subsection{Calibrating the BBA}\label{sec:calibration}

To calibrate the BBA, we perform Monte Carlo (MC) simulations by first generating fake light curves and fitting the BBA to them.  The light curves are generated by first splitting the 2.0 and 3.0 daily map data sets into declination and right ascension bands. There are four declination bands, centered on [5, 15, 25, 35] degrees, and each being 10 degrees wide.  These declination values were selected as HAWC is most sensitive to overhead events (declination=19 deg).  There are 36 right ascension bands, centered on [5, 15, ..., 355] degrees.  Next, for each declination band, 36 ROIs are defined, each centered on a right ascension band.  Each ROI has a radius of $1.44^{\circ}$.  This radius was determined by finding the median radius of the IceCube alerts used in this analysis.

With the ROIs defined, the maximum flux search is run for each ROI for each declination band.  A distribution of the maximum flux values is created by concatenating all the maximum flux values and associated uncertainties for each ROI.  Fake light curves are generated by randomly sampling the flux distribution and its associated uncertainty to create a light curve with 3000 flux values.  We used HAWC data in lieu of simulations as simulations were unable to replicate detector effects present in the daily maps.  The large number of possible observations ($\approx 100,000$) greatly reduces the probability that a flare event might be captured in constructing the fake light curves.

Once one fake light curve is generated, it is fit to the BBA with a variable value for the \emph{ncp\_prior}.  The \emph{ncp\_prior} values range from 1.0 to 7.0 (more sensitive to less sensitive), with steps of 0.25. This gives a wide range of values that can be tested quickly, as finer steps sizes greatly increase computation time.  For every \emph{ncp\_prior} value, 1000 light curves (trials) are generated and then fit.  After each run, the change point number $n_{cp}$ is stored to determine how many blocks were fit to the fake light curve.  One note is that, along with $n_{cp}$, the BBA results include several other parameters, two of which are the location and value at each change point.  These can be used to find potential flares and their significance in the HAWC dataset.  

\begin{figure*}[ht!]
    \centering
    \includegraphics[width=0.8\linewidth]{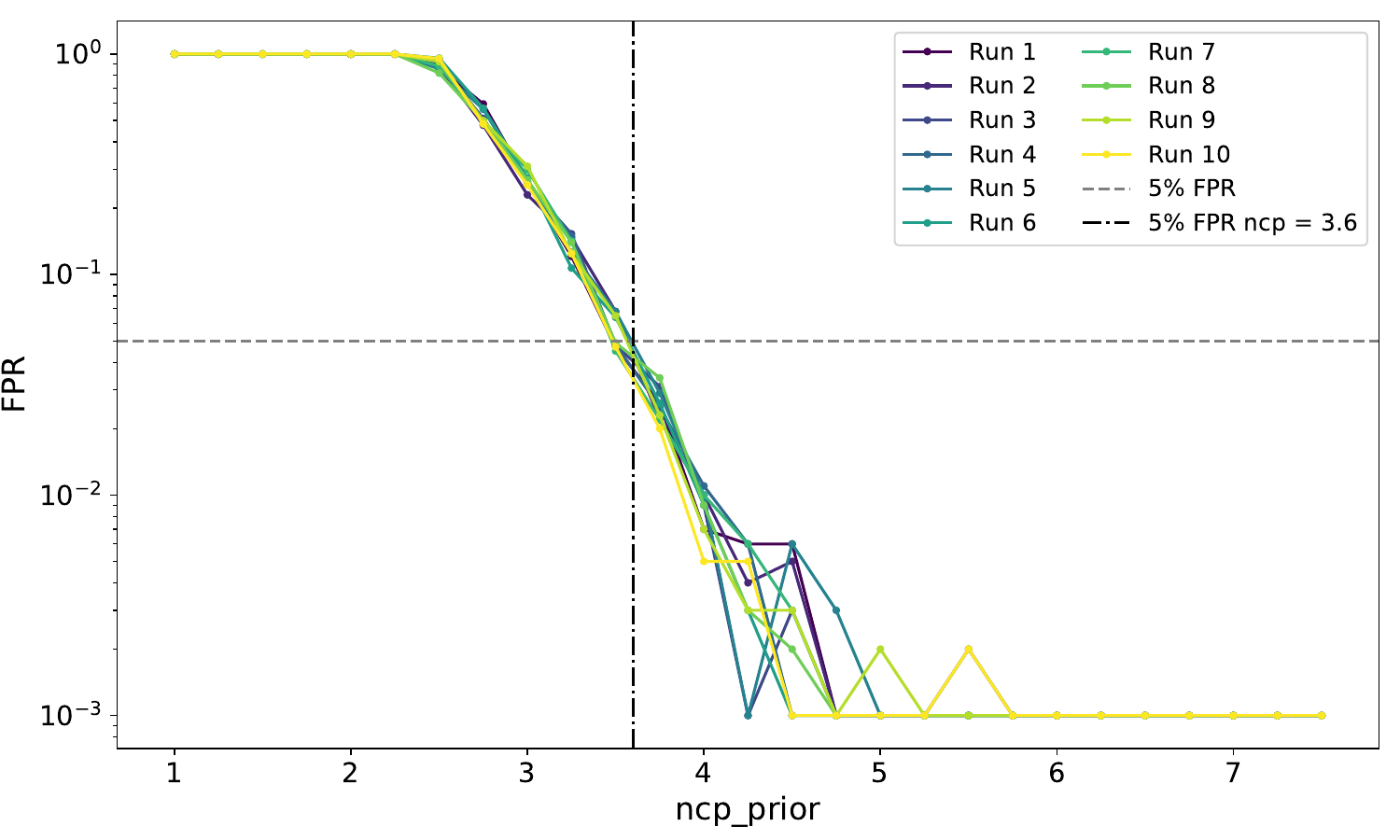}
    \includegraphics[width=0.8\linewidth]{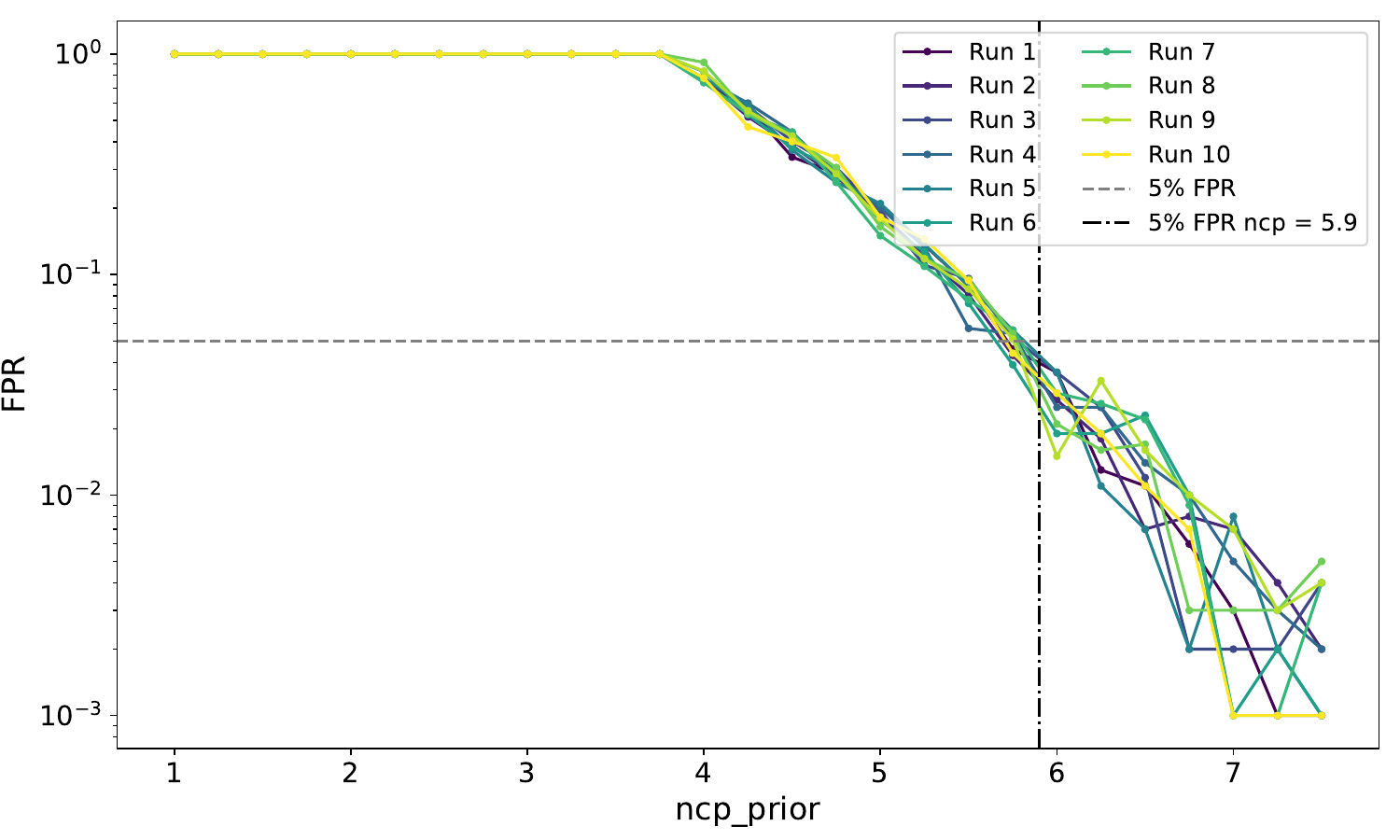}
    \caption{The results for the 10 calibrations runs for the 2.0 (upper) and 3.0 (lower) maps.  For low \emph{ncp\_prior} values, both index fits return over a 100\% FPR, which is floored to 1.  As the \emph{ncp\_prior} increases, the FPR decreases until it crosses the 0.05 ($5\%$) threshold, indicated by the dashed horizontal line.  The value where this happens is indicated by the vertical dashed line and is 3.6 for the 2.0 maps and 5.9 for the 3.0 maps. If the FPR gets below 0.001, it is set to 0.001 for readability. }
    \label{fig:calibration}
\end{figure*}

At the end of the 1000 trials, the mean $n_{cp}$ is computed to get a False Positive Rate (FPR), defined as having $n_{cp}>0$. The \emph{ncp\_prior} values are scanned to achieve a FPR of 5\%.  This process is repeated 10 times to test the stability of the 5\% FPR \emph{ncp\_prior} value.  The results of the 10 trials for both 2.0 and 3.0 maps are given in Figure \ref{fig:calibration}.  The \emph{ncp\_prior} values found using this method are 3.6 and 5.9 for the 2.0 and 3.0 maps, respectively.  

These two values indicate that the 2.0 map are more sensitive than the 3.0 map due to the smaller \emph{ncp\_prior} (see Section 2.7 in \cite{BB_paper} for details).  This is expected due to the spectral assumption made for both; the 2.0 index weighs higher energy events more significantly \citep{Crab_2024}, while the 3.0 assumption places more weight on those lower energy events. Higher energy events embed more energy into the detector, making energy reconstruction and measuring flux more accurate.  This corresponds to broadly lower uncertainties for higher energy events, making the 2.0 map more sensitive compared to the 3.0 map. A direct comparison for the two maps is discussed further in Section \ref{sec:sensitivity}.

\section{Results}\label{sec:results}

\begin{table*}
   \centering
   \begin{tabular}{c|c|c|c|c|c|c|c|c|c}
    \toprule
    \multicolumn{5}{c|}{IceCube Alert Parameters} & \multicolumn{4}{c}{HAWC Results} & \multicolumn{1}{|c}{Nearest Gamma-Ray Source} \\
    \hline
    IceCube alert & RA & DEC & Radius & $s$ & $\text{Date}_{\text{det}}$ & $\sigma_{BBA}$ & RA & DEC & Name (Distance) \\
    Name (MJD) & (deg) & (deg) & (deg) &  & MJD &  & (deg) & (deg) & (deg) \\
    \hline
      
      11-12-08B (55903) & 165.19 & 38.49 & 4.58 & 0.49 & 59066 & 20.6 & 166.16 & 38.21 &  Mrk 421 (0.8) \\
      14-06-03B (56822) & 9.71 & 7.56 & 0.71 & 0.38 & 57185 & 2.55 & 9
      49 & 7.59 & -- \\
      15-06-09B (57182) & 49.53 & 0.3 & 0.87 & 0.31 & 57328 & 1.66 & 50.1 & 0.78 & --\\
      15-09-23G (57288) & 103.23 & 3.96 & 0.80 & 0.33 & 57190 &  2.55 & 103.36 & 4.59 & -- \\
      15-10-13B (57308) & 178.72 & 52.37 & 1.12 & 0.52 & 57423 & 2.45 & 177.95 & 52.22 & -- \\
      17-10-06B (58032) & 132.63 & 17.23 & 1.34 & 0.37 & 57309 & 1.79 & 133.64 & 17.5 & -- \\
      18-06-13B (58282) & 38.06 & 11.53 & 4.79 & 0.41 & 57199 & 2.6 & 33.93 & 13.21 & 4C+13.14 (1.8) \\
      19-03-17B (58559) & 81.23 & 3.21 & 4.97 & 0.26 & 57219 & 2.87 & 78.97 & -1.12 & NVSS J052140+010257 (2.3) \\
      19-07-12B (58676) & 76.64 & 12.75 & 4.96 & 0.30 & 57199 & 2.82 & 74.53 & 12.14 & PKS 0459+135 (1.3) \\
      19-12-31B (58848) & 48.47 & 20.11 & 5.46 & 0.46 & 57089 & 3.02 & 48.69 & 15.91 & 4FGL J0312.7+2012 (.03) \\
      21-09-22G (59479) & 60.73 & -4.18 & 0.6 & 0.92 & 57184 & 2.37 & 61.35 & -3.99 & -- \\
      22-03-04G (59642) & 48.78 & 4.48 & 6.46 & 0.61 & 57168 & 3.14 & 49.04 & 5.64 & 	NGC 1218 (1.7) \\
      22-09-18B (59840) & 75.15 & 3.58 & 3.8 & 0.48 & 57089 & 3.88 & 75.15 & 7.07 & MG1 J050533+0415 (1.3) \\
      24-03-07G (60376) & 239.63 & 39.94 & 15.4 & 0.61 & 57251 & 9.42 & 253.65 & 39.69 & Mrk 501 (10.6) \\
      24-08-08B (60530) & 55.77 & 31.83 & 4.24 & 0.44 & 57218 & 3.71 & 56.87 & 31.0 & 4FGL J0341.9+3153c (0.3) \\
      26-02-19B (61090)  & 75.87 & 14.63 & 0.51 & 0.31 & 57192 & 1.94 & 75.45 & 14.32 & --

   \end{tabular}
   \caption{The IceCube alerts and location parameters that were detected with this transient search on the index 2.0 maps are given in the first five columns.  The $s$ value is the astrophysical probability for that alert. The next four columns give the highest significance change point date $\text{Date}_{\text{det}}$ and corresponding significance $\sigma_{BBA}$  found by the BBA, along with its location. Note that these are pre-trial significances.  The last column indicates whether a GeV or TeV gamma-ray source exists within the alert ROI.  These sources are from either from TeVCat \citep{TeVCat} or 4FGL \citep{4FGL}. These detections correspond to a 4.2\% coincident detection rate.}
   \label{tab:index_2_values}
\end{table*}

\begin{table*}
   \centering
   \begin{tabular}{c|c|c|c|c|c|c|c|c|c}
    \toprule
    \multicolumn{5}{c|}{IceCube Alert Parameters} & \multicolumn{4}{c}{HAWC Results} & \multicolumn{1}{|c}{Nearest Gamma-Ray Source} \\
    \hline
    IceCube alert & RA & DEC & Radius & $s$ & $\text{Date}_{\text{det}}$ & $\sigma_{BBA}$ & RA & DEC & Name (Distance) \\
    Name (MJD) & (deg) & (deg) & (deg) &  & MJD &  & (deg) & (deg) & (deg)\\
    \hline
      
    11-12-08B (55903) & 165.19 & 38.49 & 4.58 & 0.49 & 59066 & 19.31 & 166.42 & 38.11 & Mrk 421 (0.8) \\
    12-08-07G (56146) & 330.07 & 1.42 & 0.68 & 0.74 & 57013 & 1.98 & 330.21 & 2.13 & -- \\
    14-01-22B (56679) & 138.82 & 37.45 & 4.42 & 0.46 & 59036 & 3.85 & 141.81 & 41.21 & NVSS J023308+374201 (0.5) \\
    15-01-29B (57051) & 359.51 & 6.39 & 4.32 & 0.33 & 59379 & 3.04 & 1.89 & 3.88 & 4FGL J0000.5+0743 (1.5) \\
    16-03-07B (57454) & 91.32 & 10.47 & 5.73 & 0.28 & 57104 & 4.16 & 92.77 & 11.19 & 	NVSS J060835+114942 (1.6) \\
    16-04-27B (57505) & 240.29 & 9.71 & 0.48 & 0.45 & 57104 & 1.93 & 240.42 & 9.82 & -- \\
    16-07-20B (57589) & 60.25 & 29.23 & 7.7 & 0.37 & 59379 & 4.28 & 65.70 & 25.78 & 87GB 035856.9+272842 (1.7) \\
    16-11-03B (57695) & 40.87 & 12.52 & 0.84 & 0.31 & 57104 & 2.45 & 41.17 & 12.6 & GB6 J0244+1320 (0.8) \\
    18-06-13B (58282) & 38.06 & 11.53 & 4.79 & 0.41 & 57105 & 4.78 & 39.33 & 9.44 & 	4C+13.14 (1.9) \\
    20-04-10B (58949) & 242.58 & 11.61 & 8.61 & 0.30 & 59379 & 4.84 & 235.15 & 8.54 & 4C+10.45 (1.2) \\
    20-06-14B (59014) & 33.84 & 31.61 & 4.03 & 0.41 & 59379 & 4.41 & 37.49 & 31.13 & CRATES J022048+324116 (1.5) \\
    21-05-16B (59350) & 91.76 & 9.52 & 0.73 & 0.28 & 57104 & 1.99 & 91.8 & 9.29 & -- \\
    23-01-22B (59966) & 16.79 & 7.78 & 3.31 & 0.27 & 57103 & 4.31 & 14.94 & 9.03 & 	GB6 J0100+0745 (1.7) \\
    23-04-05B (60039) & 120.85 & 9.75 & 3.05 & 0.30 & 57104 & 3.62 & 121.99 & 10.62 & 	NVSS J080159+100535 (0.5) \\
    23-04-16B (60050) & 345.85 & 9.14 & 0.84 & 0.34 & 57103 & 3.69 & 345.45 & 8.69 & -- \\
    23-07-08G (60133) & 270.7 & 8.46 & 3.56 & 0.56 & 57105 & 4.08 & 272.11 & 11.68 & 4FGL J1755.9+0953 (1.2) \\
    24-08-08B (60530) & 55.77 & 31.83 & 4.24 & 0.45 & 59379 & 3.98 & 56.82 & 31.83 & 4FGL J0341.9+3153c (0.2) \\
    25-12-18B (61027) & 353.53 & 44.14 & 0.51 & 0.38 & 60108 & 2.43 & 353.31 & 44.25 & -- 

   \end{tabular}
   \caption{The IceCube alerts and location parameters that were detected with this transient search on the index 3.0 maps are given in the first five columns.  The $s$ value is the astrophysical probability for that alert.  The next four columns give the highest significance change point date $\text{Date}_{\text{det}}$ and corresponding significance $\sigma_{BBA}$  found by the BBA, along with its location. Note that these are pre-trial significances.  The last column indicates whether a GeV or TeV gamma-ray source exists within the alert ROI.  These sources are from either from TeVCat \citep{TeVCat} or 4FGL \citep{4FGL}.  These detections correspond to a 4.7\% coincident detection rate.}
   \label{tab:index_3_values}
\end{table*}

\begin{figure*}[htbp]
    \centering
    \includegraphics[width=1.0\linewidth]{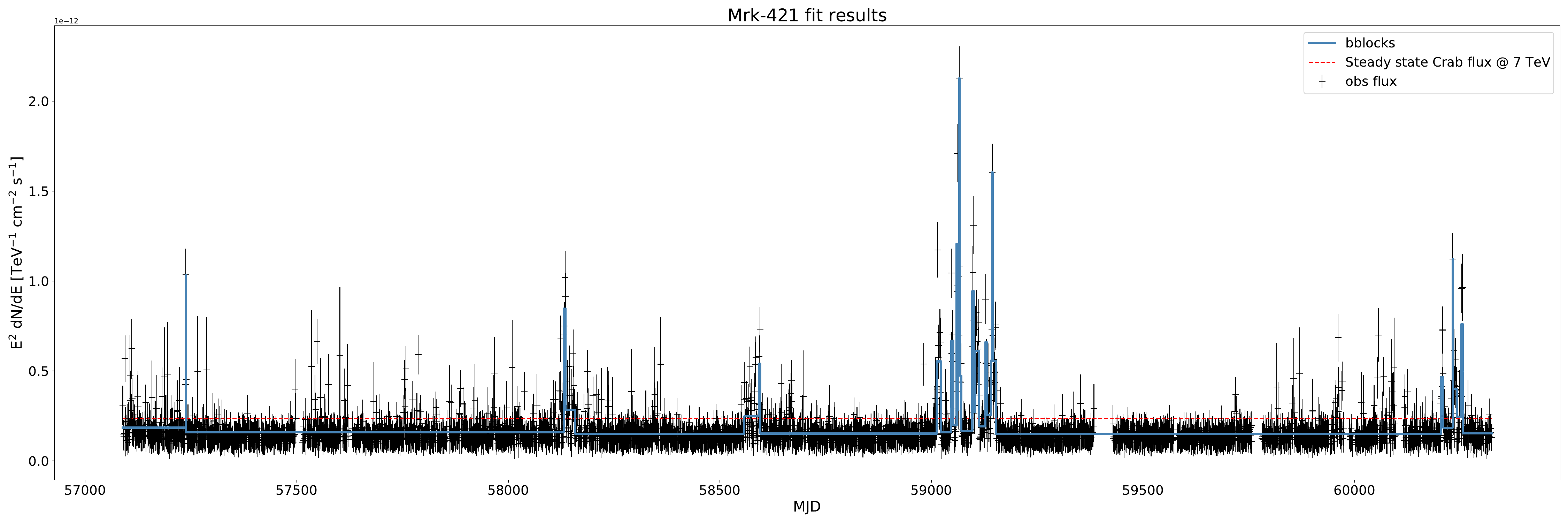}
    \includegraphics[width=1.0\linewidth]{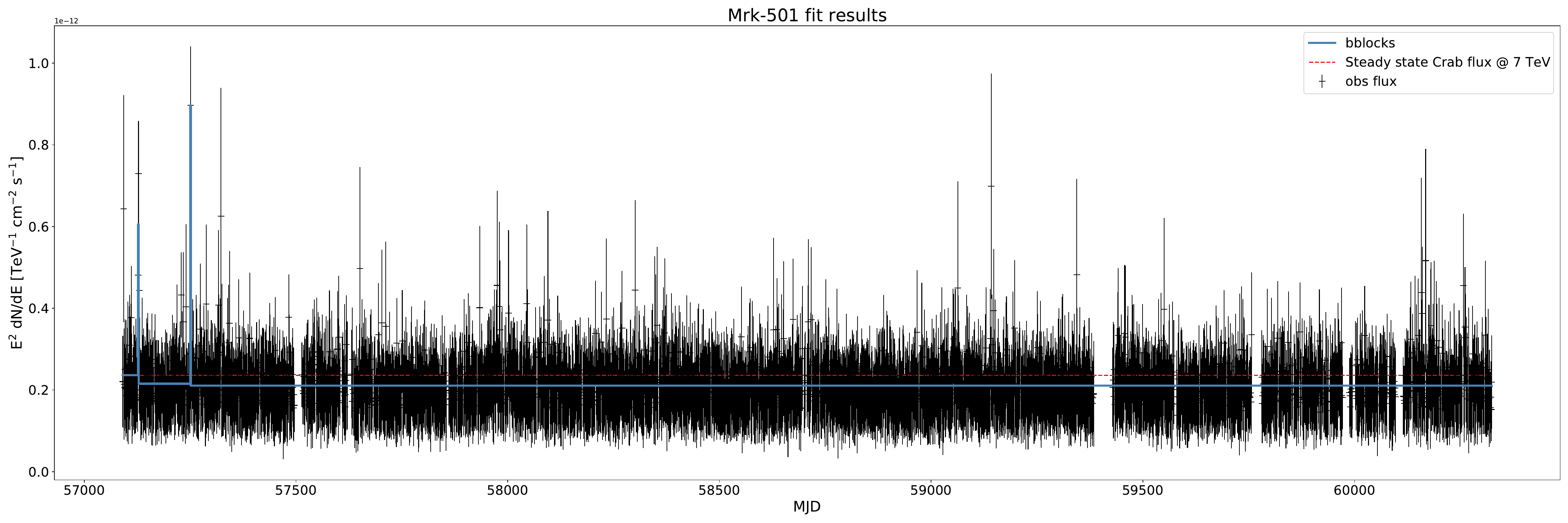}
    \caption{The light curves and BBA fits to the HAWC daily light curves for the neutrino alerts containing Mrks 421 (upper plot) and 501 (lower plot).  The alerts that captured the two Markarians are 11-12-08B and 24-03-07G, respectively. Mrk 421's alert occurred before HAWC started operations, and Mrk 501's alert occurred just after the cut-off for the daily maps. Neither IceCube alert occurred within the data set available for this analysis.}
    \label{fig:mrklc}
\end{figure*}

The results for the BBA fits to the HAWC data are presented in Tables \ref{tab:index_2_values} and \ref{tab:index_3_values} for the 2.0 and 3.0 maps, respectively.  The alerts listed in the two tables correspond to ``coincident detections,'' where the BBA finds $n_{cp}>0$ for a given alert's light curve.  These coincident detections are purely spatial in nature, with no clear temporal matching observed.  There are 16 (4.2\%) and 18 (4.7\%) spatial coincident detections for the two maps.  This is in line with expected background detection rate of 5\%.  Of the detected alerts, 3 contained HAWC steady-state $\sigma>5$ hotspots in the time-integrated map.  This is in contrast to the results found in \cite{icat1}, where 8 3HWC sources are found.  This is because these authors scanned all IceCube alerts for known sources, while here we only consider the alerts that triggered the BBA.  2 of the 3 detections correspond to the AGNs Mrk 421 and 501 (11-12-08B and 24-03-07G) and are discussed further in Section \ref{sec:multipain}. The third alert, 16-03-07B captures part of the Geminga TeV Halo \citep{3HWC} at the edge of its 5.8 degree containment radius.  Given the large alert radius and the low probability of the neutrino be astrophyisical at $s=0.28$, the alert is most likely not correlated to the TeV halo. More testing is needed to confirm the astrophysical nature of this alert and is beyond the scope of this work.

\begin{figure*}[ht!]
    \centering
    \includegraphics[width=0.95\linewidth]{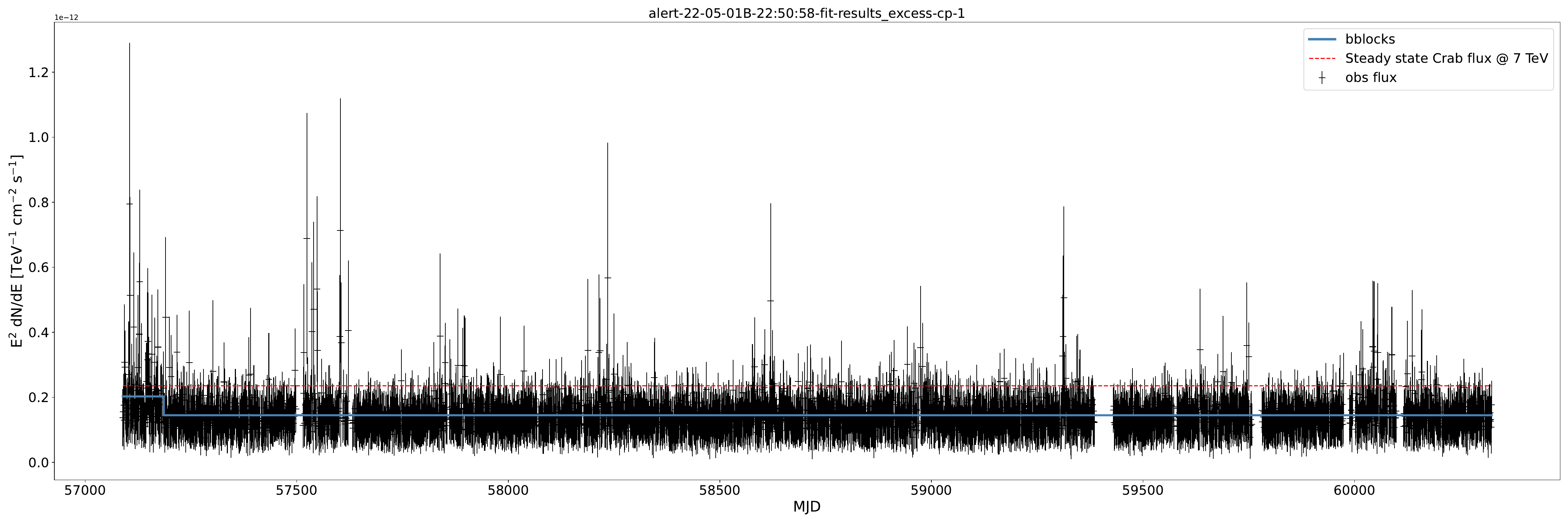}
    \includegraphics[width=0.95\linewidth]{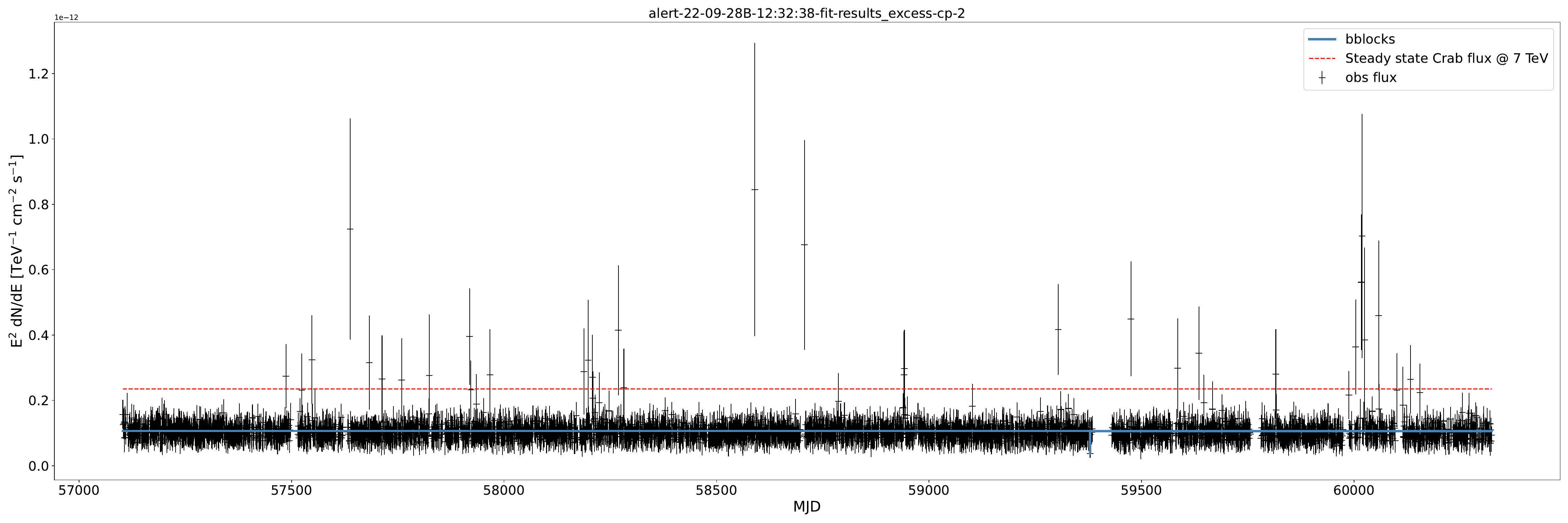}
    \caption{Examples of false positives for the 2.0 (upper plot) and 3.0 (lower plot) maps.  Aside from the two Markarians, the 2.0 maps all share this slight excess in the first year of data.  For the 3.0 maps, there's generally a low-flux value around MJD 59400, though it moves slightly, depending on the map.}
    \label{fig:fp-lightcurves}
\end{figure*}

We consider the 2.0 results first. 14 light curves have only one change point in the BBA analysis, while two alerts have greater than one change point.  These single change point alerts are all located within the first year of HAWC's operation and show a transition from higher to lower flux.  The higher-average flux duration lasts from 100 to 300 days, depending on the alert.  Given the common date range, the higher flux is most likely due to detector maintenance and calibration shortly after operations started as construction finished shortly after HAWC came online in March 2015.  This is addressed in the full time-integrated data set, but not for the daily maps, hence the discrepancy. An example is shown in the upper panel in Figure \ref{fig:fp-lightcurves}.

As for the 3.0 results, they broadly do not suffer from the same systematic issue as the 2.0 results.  Aside from two, all light curves are fit by at least two change points (three blocks) to the data.  These points are scattered across the whole date range (2015-2024), though they are being fit to under rather than the over fluctuations indicative of flares.  These points are slightly below the mean flux distribution and are well constrained.  An example is shown in the lower figure in Figure \ref{fig:fp-lightcurves}.

\subsection{Look-Elsewhere Effect}

Given the extremely large number of data points observed, there is a high likelihood of capturing $>5\sigma$ events from high background fluctuations.  This is the Look-Elsewhere Effect and is briefly addressed here.  Of the alerts, 16\% and 28\% of the index 2 and index 3 maps had pixels over $5\sigma$.  Two factors are considered when classifying an event as a potential source: the pre-trial significance being $>5\sigma$ and it being at the trigger point of the BBA.  HAWC significance maps show pre-trial significances, with the values shown in Tables \ref{tab:index_2_values} and \ref{tab:index_3_values} being such values.  To find post-trial significances, an extensive series of simulations are done to obtain the chance probability of $>5\sigma$ detections. The resulting trial correction can reduce a source's significance by roughly $0.5\sigma$ to $1\sigma$.  As such, this is only impactful if the event is near the $5\sigma$ threshold.  From Tables \ref{tab:index_2_values} and \ref{tab:index_3_values}, the only alerts for which the BBA found events $>5\sigma$ are sufficiently significant at $>9\sigma$ that post-trial studies were not necessary.  No other coincidence crosses the $5\sigma$ threshold. 

\subsection{Sensitivity Studies}\label{sec:sensitivity}

\begin{table*}[ht!]
   \centering
\begin{tabular}{c|c|c|c|c|c|c|c|c|c|c|c|c|c|c}
\toprule
\multicolumn{1}{c|}{Flare Duration} & \multicolumn{7}{c|}{Spectral Index Sensitivity} & \multicolumn{7}{|c}{Expected IceCube Observation} \\
    Intrinsic Crab Flux Fraction & 0.5 & 0.75 & 1.0 & 1.25 & 1.5 & 1.75 & 2.0 & 0.5 & 0.75 & 1.0 & 1.25 & 1.5 & 1.75 & 2.0 \\
\hline
\multicolumn{15}{c}{\textbf{Index 2.0}} \\
\hline
    1 day  & x & x & x & x & x & x & x & 0 & 0 & 0 & 0 & 0 & 0 & 0\\
    3 days & x & x & x & \checkmark & \checkmark & \checkmark & \checkmark & 0 & 1 & 1 & 1 & 2 & 2 & 2\\
    5 days & x & x & \checkmark & \checkmark & \checkmark & \checkmark & \checkmark & 1 & 1 & 2 & 2 & 3 & 4 & 4\\
    10 days & x & \checkmark & \checkmark & \checkmark & \checkmark & \checkmark & \checkmark & 2 & 3 & 4 & 5 & 7 & 8 & 9\\
\hline
\multicolumn{15}{c}{\textbf{Index 3.0}} \\
\hline
    1 day  & x & x & x & x & x & x & x & 0 & 0 & 0 & 0 & 0 & 0 & 0\\
    3 days & x & x & x & x & \checkmark & \checkmark & \checkmark & 0 & 0 & 0 & 0 & 0 & 0 & 1\\
    5 days & x & x & x & \checkmark & \checkmark & \checkmark & \checkmark & 0 & 0 & 0 & 1 & 1 & 1 & 1\\
    10 days & x & x & \checkmark & \checkmark & \checkmark & \checkmark & \checkmark & 0 & 1 & 1 & 2 & 2 & 3 & 3\\
\hline
\end{tabular}
\caption{Detection rate of the BBA for the index 2 and index 3 data sets for different flux levels of intrinsic Crab-like flares, and the corresponding number of neutrino events theoretically observed by IceCube. Broadly, both observatories need multi-day Crab-like flux levels for a given event before they are detected.  For the 2.0 map, HAWC needs a 3-day flare at 1.25 times the Crab while the 3.0 map needs a 1.5 times the Crab for 3 days.  The IceCube neutrino counts follow a similar trend.  These results are for the 18 degree event, with the -12 and 49.3 degree events yielding slightly less sensitive results.}
\label{tab:sensitivity_study_results_combined}
\end{table*}

\begin{figure*}[htbp]
    \centering
    \includegraphics[width=1.0\linewidth]{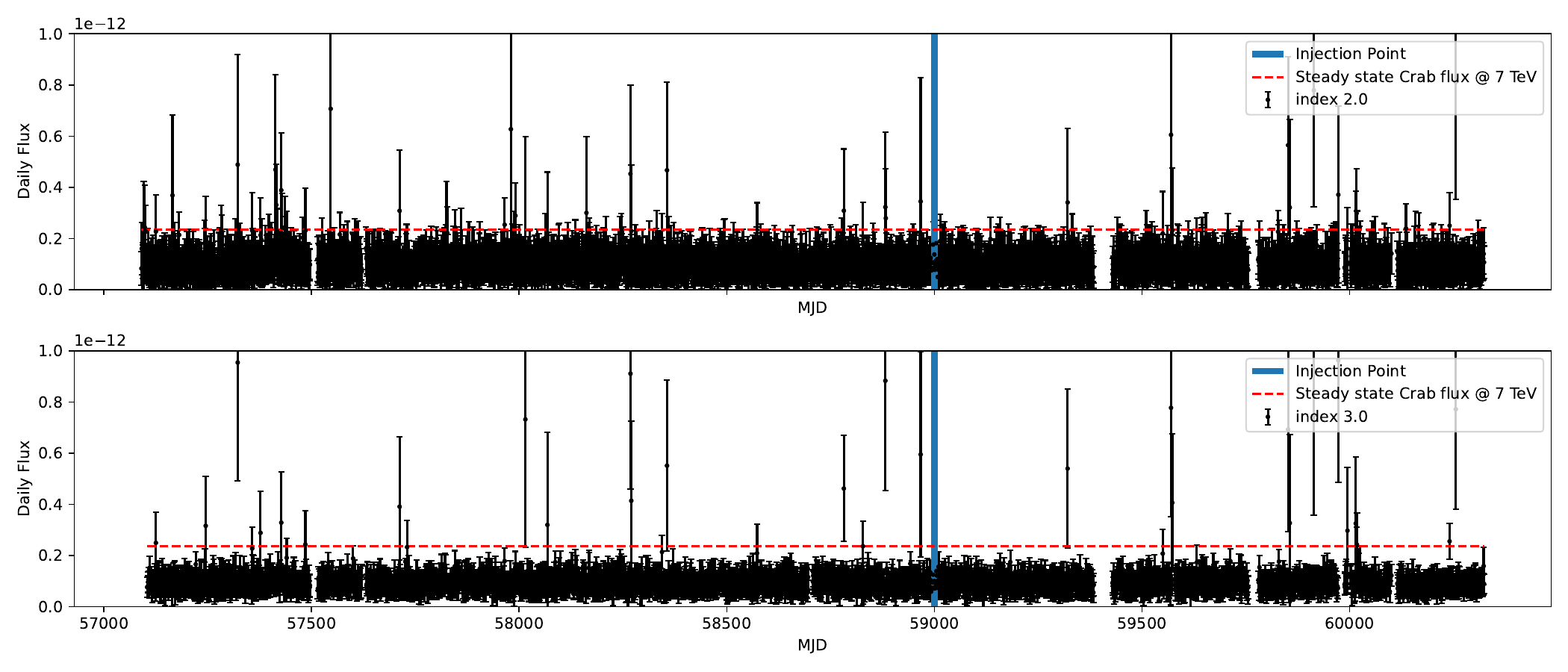}
    \caption{Comparing the 2.0 (upper plot) and 3.0 (lower plot) data sets for the IceCube alert 24-04-24G used in the injection studies.  The simulated Crab events are injected at MJD 59000, indicated by the vertical blue line. }
    \label{fig:2_3-example}
\end{figure*}

To test the sensitivity of the BBA, we perform a series of injection tests using the gamma-ray light curves generated at the location of IceCube alerts.  We test three different alerts, each at different declinations.  These are at a high declination of 49.3 deg (24-07-25B), a near zenith alert at 18.6 degrees (22-05-01B), and a low declination alert at -12 deg (20-11-30G).  These alerts were chosen due to their wide declination spread, they did not trigger the BBA, and there is no significant variation from the baseline flux.  The high-declination alert for both index maps is shown in Figure \ref{fig:2_3-example}. 

After selecting the light curve, we injected fake flares of varying fluxes (in Crab units) with durations of 1, 3, 5, and 10 days.  The source is injected at date MJD 59000 (May 31, 2020).  The flux-scaling factor ranges from 0.5 to 2.0, stepping by 0.25. To determine what a reasonable value for the Crab would be for a given flare, we performed the max pixel search described in Section \ref{sec:data_collect} on the Crab for both index maps before saving the results to disk.  We then randomly sampled this data for the needed number of days (1,3,5,10) before applying the desired scaling factor and injecting them into the three light curves at MJD 59000.  This process was done 1000 times for each length, flux scaling, and index.

The results for the 18 deg declination simulations are shown on the left side of Table \ref{tab:sensitivity_study_results_combined}, with the off-declination simulations yielding slightly less sensitive results.  Broadly, the algorithm needs at least a flare with Crab-like flux lasting multiple days at higher ($>10$ TeV) energies for the BBA to detect it.  It should be noted that it does not need to be a constant flare, only that its flux reaches Crab-like levels for a given day.  For lower energies, the algorithm is less sensitive and needs greater than a Crab-like flare to be detected.  Neither index map is sensitive to a single day flare if it is less than twice the Crab's flux.  

\subsection{Neutrino Rate Estimation}

In addition to HAWC's sensitivity to potential flares, we estimate the number of neutrinos that IceCube's alert system would expect to see.  To do this, we use the effective area values provided from IceCat 1 \citep{icat1} and the following relation from \cite{alison_thesis}:

\begin{equation}
    N = \int A_{eff}^{alert}(E,\delta) \times \Phi(E) dE.
\end{equation}

\noindent This integrates the product of the effective area $A_{eff}^{alert}$, which is a function of energy and the declination of the flare $\delta$ and the energy spectrum of the neutrino source \cite{icat1}. This produces the number of expected neutrinos.  We assume a neutrino spectrum consistent with the same Crab-like events tested above. Lastly, we only consider the intrinsic Crab data and do not apply an EBL model.

The results of this study are shown on the right side of Table \ref{tab:sensitivity_study_results_combined}.  Broadly speaking, IceCube's public alert neutrino rate is similar to HAWC's sensitivity, where IceCube would release a neutrino alert roughly when HAWC triggers on a potential flare. 

\section{Multi-messenger Analysis with IceCube Alerts}\label{sec:multipain}

\begin{table*}
   \centering
   \begin{tabular}{c|c|c}
      \toprule
       & Mrk 421 & Mrk 501 \\
      \hline
      $\sqrt{TS}$ & 121 & 26 \\
      RA (deg) & $166.14 \pm 0.03$ & $253.54 \pm 0.01$ \\
      DEC (deg) & $38.19 \pm 0.02$ & $39.76 \pm 0.01$ \\
      $N_0 [10^{-12}$ TeV$^{-1}$ cm$^{-1}$ s$^{-1}$] & $6.85 \pm 0.23$ & $1.07 \pm 0.05$\\
      $\alpha$ & $-2.437 \pm 0.027$ & $-2.57 \pm 0.05$ \\
      $E_c$(TeV) & $9.9 \pm 1.2$ & --\\
      $E_{max}$ & 16.3 & 15.0
   \end{tabular}
   \caption{Fit parameters for the two Markarians.  The fits assumed a pivot of 2 TeV.  The max energy is presented at the $2\sigma$ threshold. Additionally, the uncertainties are statistical only.}
   \label{tab:mrk_fits}
\end{table*}

\begin{figure*}[htbp]
    \centering
    \includegraphics[width=0.49\linewidth]{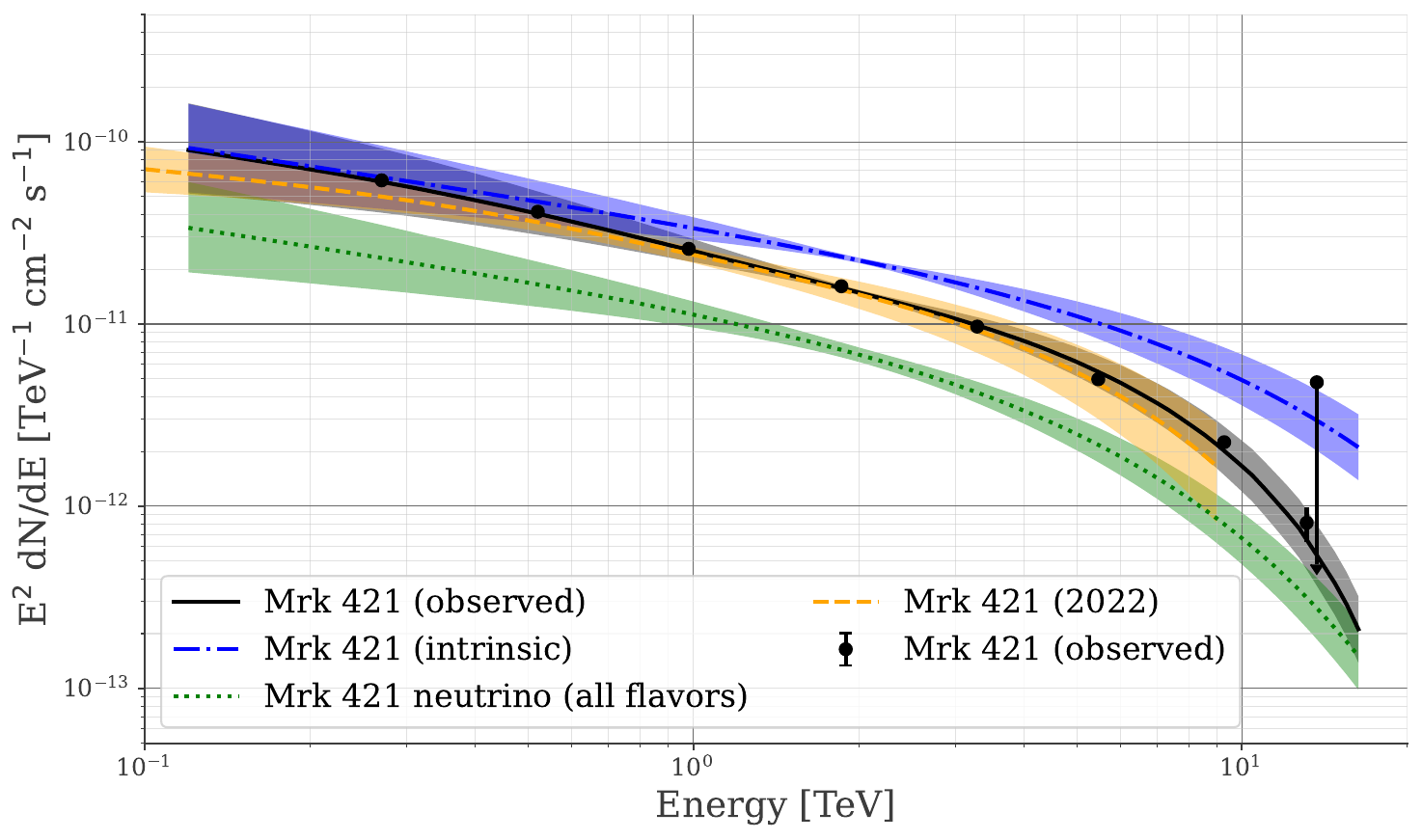}
    \includegraphics[width=0.49\linewidth]{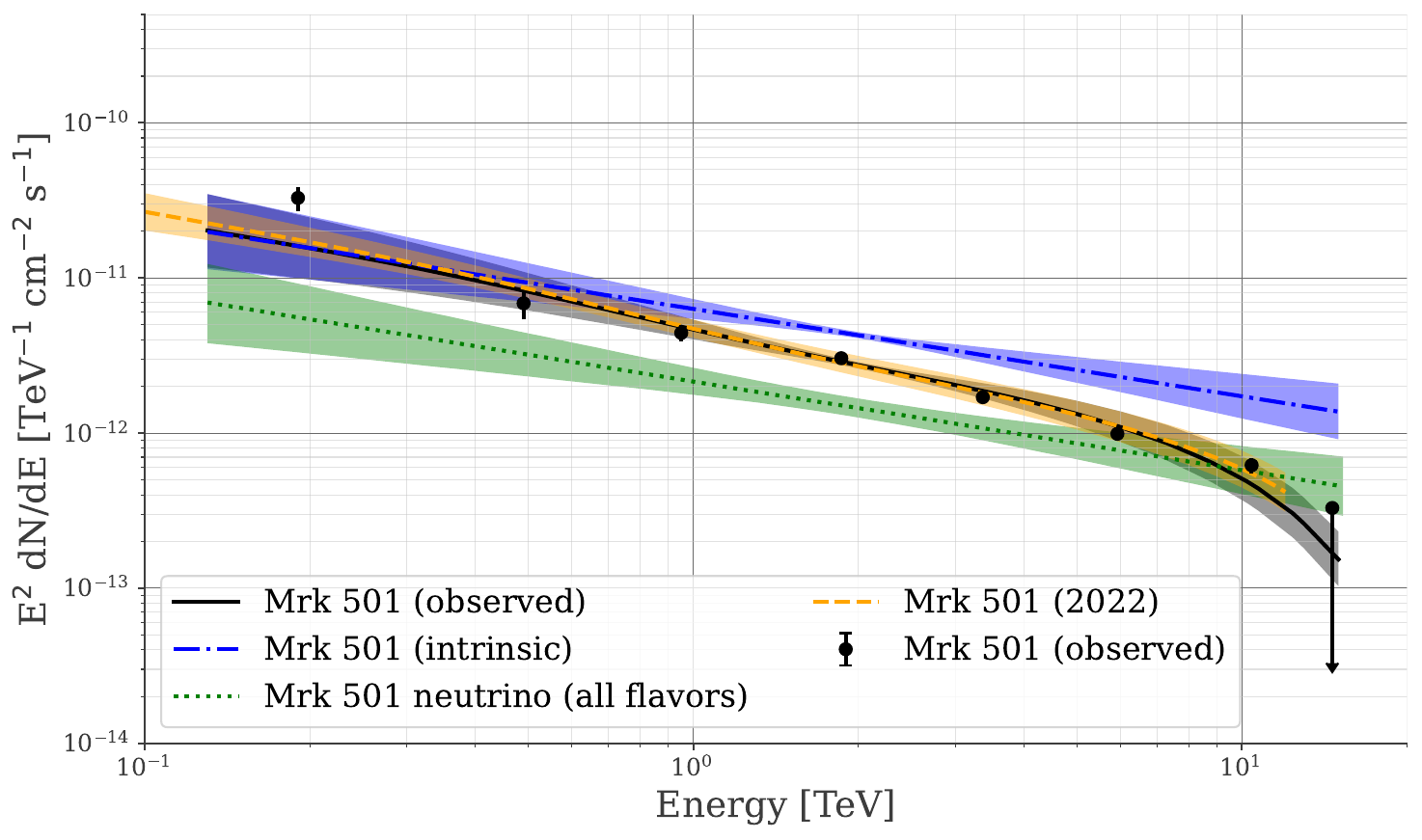}
    \caption{The spectral-energy distributions for Mrk 421 (right) and Mrk 501 (left).  The fits for this work are solid black with their intrinsic spectra indicated in dash dotted blue.  The orange dashed lines are the previous results from \cite{hawc_mrk_steadystate}.  Lastly, the neutrino spectra from Section \ref{sec:neutrino_stuff} are shown in green. One note is that Mrk 501 is modeled as a power law, but EBL absorption causes a cut-off to develop at approximately 10 TeV.}
    \label{fig:mrkseds}
\end{figure*}

As AGNs have been proven to emit neutrinos \citep{TEXAS} and two of the neutrino alerts include Mrks 421 (11-12-08B) and 501 (24-03-07G) inside their 90\% containment radii, we perform an analysis to determine if the Markarians could be the source of the neutrino alerts.  Both Markarians were quiescent during these events, so a steady-state emission model is assumed.

While HAWC does have a long-term analysis of the two Markarians \citep{hawc_mrk_steadystate}, that analysis only used approximately 1000 days of data.  For this analysis, we use the 2565 day time-integrated data set.  Additionally, we used the neural-network energy estimator introduced in Section \ref{sec:datasets}.  This data uses a superior background rejection algorithm that is discussed in \cite{hawcML}.  

To fit the two Markarians, the spectral models from \cite{hawc_mrk_steadystate} are used.  These are a cut-off power law and a power law for Mrks 421 and 501 respectively.  Additionally, the Extragalactic Background Light (EBL) must be considered.  The EBL absorbs high energy gamma-rays and re-emits them at lower energies.  The spectral models fitted for the Markarians are

\begin{equation}
    \frac{dE}{dN} = N_o \left( \frac{E}{E_p} \right) ^{\alpha} \times \exp(-\tau(E,z))
\end{equation}

\noindent and

\begin{equation}
    \frac{dE}{dN} = N_o \left( \frac{E}{E_p}\right)^{\alpha} \times \exp(\frac{-E}{E_c}) \times \exp(-\tau(E,z)).
\end{equation}

The pivot energy is fixed to 2 TeV and the EBL model is given by $\tau(E,z)$.  It is a function of the energy and redshift. The model considered is by \cite{franceschini2008extragalactic} with redshifts of 0.03 and 0.031 for Mrks 421 and 501, respectively \citep{hawc_mrk_steadystate}.  We fit the models using the Multi-Mission Maximum Likelihood (3ML) framework \cite{3ML}\footnote{\url{https://github.com/threeML/threeML}} with the HAWC Accelerated Likelihood (HAL) plugin from \cite{hal}\footnote{\url{https://github.com/threeML/hawc_hal}}. The results of these fits are given in Table \ref{tab:mrk_fits} and their spectra are shown in Figure \ref{fig:mrkseds}.  Additionally, their predicted neutrino spectra (discussed below) are also presented.

\subsection{Neutrino Emission}\label{sec:neutrino_stuff}

To test for potential neutrino emission, the HAWC results of Mrks 421 and 501 are converted to neutrino spectra.  To do this, the following relation from \cite{neutrino_spectum_convert} is used: 

\begin{equation}
    E_{\gamma} F_\gamma(E_\gamma) = e^{\frac{-d}{\lambda_{\gamma \gamma}}} \frac{2}{3K}\sum_{\nu_\alpha} E_\nu F_{\nu_\alpha}(E_\nu).
    \label{eq:spect_convert}
\end{equation}

\noindent This converts the observed gamma-ray flux and energies ($ F_\gamma, E_\gamma$) to a predicted neutrino spectrum with flux and energy ($ F_\nu, E_\nu$) by summing over the 3 neutrino flavors (tau, electron, muon).  Absorption from the EBL is considered with the distance $d$ and the attention length $\lambda_{\gamma \gamma}$. The $K$ parameter is defined by the production mechanism responsible for the observed gamma rays.  The two considered are proton-proton ($pp$) and proton-photon ($p\gamma$).  Both rely on neutral and charged pion decay to produce gamma rays and neutrinos respectively, with the key difference being the ratio of charged to neutral pions being different.  For $pp$, there are three products produced: positive, negative, and neutral pions, while $p\gamma$ interactions produce pions via $\Delta(1232)^+$ decays that either produce positive or neutral pions \cite{pdg}. Therefore, $pp$ has a charged to neutral production ratio of 2:1 while $p\gamma$ has a 1:1 ratio.  $K$ is either 2 ($pp$) or 1 ($p\gamma$).  As discussed in \cite{texas_multi} and \cite{aharonianbook}, $p\gamma$ production is favored in AGN, so $K=1$ is considered for Equation \ref{eq:spect_convert}.  

Another consideration is the relationship between produced neutrinos and gamma rays.  Each neutral pion produces 2 photons at half the pion's energy while charged pions produced an electron or positron and 3 neutrinos, each carrying roughly a quarter of the initial energy.  Therefore, the photon energy is twice that of a given neutrino.  Combining this relation, $K=1$, and a cut-off power law spectral model, a gamma-ray source like Mrk 421 can have its intrinsic ($d=0$) photon spectrum converted to a neutrino source with a spectrum given by

\begin{equation}
    \frac{dN}{dE_\nu} = 2^{\alpha} N_{0,\gamma} \left(\frac{E_\nu}{E_p}\right)^{\alpha}\times \exp \left(\frac{2E_\nu} {E_{c,\gamma}}\right).
\end{equation}

The expression for a power law spectrum removes the cut-off parameter but is otherwise unchanged.  The predicted neutrino spectra for Mrks 421 and 501 are shown in green in Figure \ref{fig:mrkseds}. Considering Mrk 421 first, the observed neutrino energy of 123 TeV appears unlikely to be produced by a steady-state emission model due to the low neutrino cut-off energy of $\approx 5$ TeV.  Meanwhile, Mrk 501 theoretically could produce a 191 TeV neutrino as it is observed by a power law with no cut-off with HAWC.  Regardless, neither source is observed at high energies to determine they could potentially produce these neutrinos.  

A final consideration is that both Markarians have been well modeled using synchrotron self-Compton (SSC) scattering using radio to TeV data previously \citep{hawc_mrk_steadystate}.  While this model assumes emission is dominated by leptonic processes, it is natural that protons would be accelerated as well.  If they achieve high enough energies, they can produce TeV-PeV neutrinos through proton-photon processes inside the jet.  Discussed extensively in Section 3 of \cite{texas_multi}, many models can produce TeV neutrinos while suppressing the corresponding photons.  To properly distinguish among these models, extensive multi-wavelength fitting is needed and is beyond the scope of this work.  The goal of the results presented here is to demonstrate whether HAWC's observed data is compatible with possible $>100$ TeV neutrino alerts.

\section{Systematic Uncertainties}\label{sec:discussion}

Considering the HAWC-IceCube coincident search first, there are three factors that affect the results from the BBA: the joint effect of calibration of the BBA and potential declination effects, and a potentially low signal/noise ratio.  As the calibration and declination dependence are correlated, they will be considered first.  The method used for calibrating the BBA only considered four of the possible eight declinations bands available.  These declination bands, centered on [5, 15, 25, 35] degrees, represent the region nearest to HAWC's zenith of 19 degrees.  As shown in Figure 3 in \cite{Crab_2024}, HAWC's PSF worsens the further off zenith the events are.  This effect is reflected in the average higher flux and larger uncertainty values that can be seen in Mrk 501's light curve in Figure \ref{fig:mrklc}.  Broadly speaking, alerts closer to zenith have a more-constrained uncertainty distribution.

Several methods were tested to address this issue.  First, the overall declination range was set to [-20, 60] degrees to help reduce this systematic.  Then, several configurations for declination bands were tested.  These ranged from 8 bands, each 10 degrees wide, to the 3 bands defined in \cite{Crab_2024}: [0, 26], [26, 37], [37, 46] degrees off zenith.  While calibrating these different configurations would succeed, applying them to real alerts usually resulted in unreasonably high detection rates of $>10\%$.  A compromise was made to focus on where HAWC is most sensitive while still being able to see a high zenith flare, if it was sufficiently strong.  This is the primary reason why most of the detections presented in Tables \ref{tab:index_2_values} and \ref{tab:index_3_values} are relatively close to HAWC's zenith.

It is also worth noting is that, while the BBA is relatively agnostic to \emph{ncp\_prior} values for moderate to high signal-to-noise ratios, it can have issues in low signal-to-noise environments \citep{BB_paper}.  This partially motivated the decision to restrict calibration to HAWC's core declination bands.  This maximizes the possibility of finding flares within 20 degrees of HAWC's zenith, but penalizes the probability of finding high-declination events.  Broadly, this analysis lies in the low signal-to-noise regime, and so the BBA calibration must be extremely thorough. Additional calibration methods may yield a more sensitive BBA, though that is beyond the scope of this work.

\section{Conclusion}\label{sec:conclusion}

In this analysis, we performed an investigation of public neutrino alerts from IceCube using HAWC.  We searched both our 2565-day steady-state and $\approx3000$ daily map data sets for potential coincidences. These consider either known gamma-ray sources or potential transient events.  A Bayesian Block Algorithm was calibrated to fit the maximum flux values collected from the daily maps using regions of interest defined by the 90\% containment radius of the IceCube alerts.  Two spectral assumptions, 2.0 and 3.0, were used to probe different energy ranges.  These achieved 4.2\% and 4.7\% coincident detection rate respectively, which is in line with a targeted false positive rate of 5\%.  We also performed injection studies to determine how sensitive HAWC is to transient events and how many neutrino alerts IceCube would release for such events. Both HAWC or IceCube need Crab-like or brighter events lasting several days to be detected or send a neutrino alert. It is hoped that larger and more powerful observatories like LHAASO and the planned Southern Wide-field Gamma-ray Observatory (SWGO) can significantly improve this search.

Two of these alerts included the active galactic nuclei Markarians 421 and 501. We investigated whether the Markarians could have produced $>100$ TeV neutrinos by fitting the two sources with HAWC's time-integrated data set.  The results are inconclusive due to the small observed energy range of the Markarians and a lack of observed flaring in conjunction with the neutrino alerts.  With new neutrino observatories like KM3net and P-ONE coming online, neutrino transient events like the IceCube alerts will hopefully become more constrained and point to more joint neutrino-gamma ray events.

\begin{acknowledgments}

\end{acknowledgments}

We acknowledge the support from: the US National Science Foundation (NSF); the US Department of Energy Office of High-Energy Physics; the Laboratory Directed Research and Development (LDRD) program of Los Alamos National Laboratory; Consejo Nacional de Ciencia y Tecnolog\'{\i}a (CONACyT), M\'exico, grants LNC-2023-117, 271051, 232656, 260378, 179588, 254964, 258865, 243290, 132197, A1-S-46288, A1-S-22784, CF-2023-I-645, CBF2023-2024-1630, c\'atedras 873, 1563, 341, 323, Red HAWC, M\'exico; DGAPA-UNAM grants IG101323, IG100726, IN111716-3, IN111419, IA102019, IN106521, IN114924, IN110521, IN102223; VIEP-BUAP; PIFI 2012, 2013, PROFOCIE 2014, 2015; the University of Wisconsin Alumni Research Foundation; the Institute of Geophysics, Planetary Physics, and Signatures at Los Alamos National Laboratory; Polish Science Centre grant, 2024/53/B/ST9/02671; Coordinaci\'on de la Investigaci\'on Cient\'{\i}fica de la Universidad Michoacana; Royal Society---Newton Advanced Fellowship 180385; Gobierno de Espa\~na and European Union-NextGenerationEU, grant CNS2023-144099; The Program Management Unit for Human Resources \& Institutional Development, Research and Innovation, NXPO (grant number B16F630069); Coordinaci\'on General Acad\'emica e Innovaci\'on (CGAI-UdeG), PRODEP-SEP UDG-CA-499; Institute of Cosmic Ray Research (ICRR), University of Tokyo. H.M. acknowledges support under grant number CBF2023-2024-1630. H.F. acknowledges support by NASA under award number 80GSFC21M0002. C.R. acknowledges support from National Research Foundation of Korea (RS-2023-00280210). We also acknowledge the significant contributions over many years of Stefan Westerhoff, Gaurang Yodh and Arnulfo Zepeda Dom\'{\i}nguez, all deceased members of the HAWC collaboration. Thanks to Scott Delay, Luciano D\'{\i}az and Eduardo Murrieta for technical support.

\bibliography{sample701}{}

@article{icecube-hardware,
  title={The IceCube Neutrino Observatory: instrumentation and online systems},
  author={Aartsen, Mark G and Ackermann, M and Adams, J and Aguilar, JA and Ahlers, M and Ahrens, M and Altmann, D and Andeen, K and Anderson, T and Ansseau, I and others},
  journal={Journal of Instrumentation},
  volume={12},
  number={03},
  pages={P03012},
  year={2017},
  publisher={IOP Publishing}
}

@article{3HWC,
  title={3HWC: the third HAWC catalog of very-high-energy gamma-ray sources},
  author={Albert, A and Alfaro, R and Alvarez, C and Camacho, JR Angeles and Arteaga-Vel{\'a}zquez, JC and Arunbabu, KP and Rojas, D Avila and Solares, HA Ayala and Baghmanyan, V and Belmont-Moreno, E and others},
  journal={The Astrophysical Journal},
  volume={905},
  number={1},
  pages={76},
  year={2020},
  publisher={IOP Publishing}
}

@article{pdg,
  title={Review of particle physics},
  author={Beringer, Juerg and Arguin, J-F and Barnett, RM and Copic, K and Dahl, O and Groom, DE and Lin, C-J and Lys, J and Murayama, H and Wohl, CG and others},
  journal={Physical Review D—Particles, Fields, Gravitation, and Cosmology},
  volume={86},
  number={1},
  pages={010001},
  year={2012},
  publisher={APS}
}

@article{neutrino_spectum_convert,
  title={Probing the Galactic origin of the IceCube excess with gamma rays},
  author={Ahlers, Markus and Murase, Kohta},
  journal={Physical Review D},
  volume={90},
  number={2},
  pages={023010},
  year={2014},
  publisher={APS}
}

@article{gamma_delayed_emission,
  title={Upstream neutrino production and delayed jet emission in the blazar GB6 J1542+ 6129},
  author={Kun, Emma and Bartos, Imre and Hadi, Breshna and G{\"o}bly{\"o}s, Anna and Tjus, Julia Becker and Biermann, Peter L and Franckowiak, Anna and Halzen, Francis and del Palacio, Santiago and Ricci, Claudio},
  journal={arXiv preprint arXiv:2605.00785},
  year={2026}
}

@book{alison_thesis,
  title={Transient Searches with the HAWC Gamma-Ray Observatory},
  author={Peisker, Alison},
  year={2021},
  publisher={Michigan State University}
}

@article{neutrino_delayed_emission,
  title={Search for neutrino emission from blazar $$\backslash$gamma $-ray flares accounting for possible neutrino time delays},
  author={Podlesnyi, Egor and Oikonomou, Foteini},
  journal={arXiv preprint arXiv:2511.01361},
  year={2025}
}

@article{texas_multi,
  title={A multimessenger picture of the flaring blazar TXS 0506+ 056: Implications for high-energy neutrino emission and cosmic-ray acceleration},
  author={Keivani, A and Murase, K and Petropoulou, M and Fox, DB and Cenko, SB and Chaty, S and Coleiro, A and DeLaunay, JJ and Dimitrakoudis, S and Evans, PA and others},
  journal={The Astrophysical Journal},
  volume={864},
  number={1},
  pages={84},
  year={2018},
  publisher={The American Astronomical Society}
}

@ARTICLE{TeVCat,
author = {{Wakely}, S.~P. and {Horan}, D.},
title = "{TeVCat: An online catalog for Very High Energy Gamma-Ray Astronomy}",
journal = {International Cosmic Ray Conference},
year = 2008,
volume = 3,
pages = {1341-1344},
adsurl = {http://adsabs.harvard.edu/abs/2008ICRC....3.1341W}
}

@article{4FGL,
  title={Fermi Large Area Telescope fourth source catalog data release 4 (4FGL-DR4)},
  author={Ballet, Jean and Bruel, P and Burnett, TH and Lott, B and Fermi-LAT Collaboration and others},
  journal={arXiv preprint arXiv:2307.12546},
  year={2023}
}

@article{GW_example,
  title={Observation of gravitational waves from a binary black hole merger},
  author={Abbott, Benjamin P and Abbott, Richard and Abbott, Thomas D and Abernathy, Matthew R and Acernese, Fausto and Ackley, Kendall and Adams, Carl and Adams, Thomas and Addesso, Paolo and Adhikari, Rana X and others},
  journal={Physical review letters},
  volume={116},
  number={6},
  pages={061102},
  year={2016},
  publisher={APS}
}

@article{ngc_1068,
  title={Evidence for neutrino emission from the nearby active galaxy NGC 1068},
  author={Abbasi, R and Ackermann, M and Adams, J and Aguilar, JA and Ahlers, M and Ahrens, M and Alameddine, JM and Alispach, C and Alves Jr, AA and others},
  journal={Science},
  volume={378},
  number={6619},
  pages={538--543},
  year={2022},
  publisher={American Association for the Advancement of Science}
}

@article{hawcradio,
  title={Longtime Monitoring of TeV Radio Galaxies with HAWC},
  author={Alfaro, R and Alvarez, C and Anita-Rangel, E and Arteaga-Vel{\'a}zquez, JC and Rojas, D Avila and Solares, HA and Babu, R and Bangale, P and Belmont-Moreno, E and Bernal, A and others},
  journal={arXiv preprint arXiv:2506.16031},
  year={2025}
}

@article{hawc_monitoring,
  title={The HAWC real-time flare monitor for rapid detection of transient events},
  author={Abeysekara, AU and Alfaro, R and Alvarez, C and {\'A}lvarez, JD and Arceo, R and Arteaga-Vel{\'a}zquez, JC and Rojas, D Avila and Solares, HA Ayala and Barber, AS and Bautista-Elivar, N and others},
  journal={The Astrophysical Journal},
  volume={843},
  number={2},
  pages={116},
  year={2017},
  publisher={IOP Publishing}
}

@article{lhaaso,
  title={The large high altitude air shower observatory (LHAASO) science book (2021 Edition)},
  author={Cao, Zhen and della Volpe, D and Liu, Siming and Bi, Xiaojun and Chen, Yang and Piazzoli, BD'Ettorre and Feng, Li and Jia, Huanyu and Li, Zhuo and Ma, Xinhua and others},
  journal={arXiv preprint arXiv:1905.02773},
  year={2019}
}

@article{IC_galactic,
  title={Observation of high-energy neutrinos from the Galactic plane},
  author={Abbasi, R and Ackermann, M and Adams, J and Aguilar, JA and Ahlers, M and Ahrens, M and Alameddine, JM and Alves Jr, AA and Amin, NM and others},
  journal={Science},
  volume={380},
  number={6652},
  pages={1338--1343},
  year={2023},
  publisher={American Association for the Advancement of Science}
}

@article{hawc_nim,
  title={The high-altitude water Cherenkov (HAWC) observatory in M{\'e}xico: The primary detector},
  author={Abeysekara, AU and Albert, A and Alfaro, R and Alvarez, C and {\'A}lvarez, JD and Araya, M and Arteaga-Vel{\'a}zquez, JC and Arunbabu, KP and Rojas, D Avila and Solares, HA Ayala and others},
  journal={Nuclear Instruments and Methods in Physics Research Section A: Accelerators, Spectrometers, Detectors and Associated Equipment},
  volume={1052},
  pages={168253},
  year={2023},
  publisher={Elsevier}
}

@article{swgo,
  title={The southern wide-field Gamma-ray observatory (SWGO): A next-generation ground-based survey instrument for VHE Gamma-ray astronomy},
  author={Abreu, P and Albert, A and Alfaro, R and Alvarez, C and Arceo, R and Assis, P and Barao, F and Bazo, J and Beacom, JF and Bellido, J and others},
  journal={arXiv preprint arXiv:1907.07737},
  year={2019}
}

@book{aharonianbook,
  title={Very high energy cosmic gamma radiation: a crucial window on the extreme Universe},
  author={Aharonian, Felix A},
  year={2004},
  publisher={World Scientific}
}

@article{TEXAS_og,
  title={Multimessenger observations of a flaring blazar coincident with high-energy neutrino IceCube-170922A},
  author={Aartsen, Mark G  and M Ackermann and others},
  journal={Science},
  volume={361},
  number={6398},
  pages={eaat1378},
  year={2018},
  publisher={American Association for the Advancement of Science}
}

@article{TEXAS,
  title={TXS 0506+ 056 with Updated IceCube Data},
  author={Abbasi, Rasha and Ackermann, Markus and Adams, Jenni and Agarwalla, Sanjib Kumar and Aguilar, Juanan and Ahlers, Markus and Alameddine, Jean-Marco and Amin, Najia\_Moureen Binte and Andeen, Karen and Anton, Gisela and others},
  year={2023},
  publisher={Sissa Medialab},
  journal={Proceedings of Science}
}

@article{icat1,
  title={IceCat-1: the IceCube event catalog of alert tracks},
  author={Abbasi, R and Ackermann, M and Adams, J and Agarwalla, S\_K and Aguilar, J\_A and Ahlers, M and Alameddine, J\_M and Amin, N\_M and Andeen, K and Anton, G and others},
  journal={The Astrophysical Journal Supplement Series},
  volume={269},
  number={1},
  pages={25},
  year={2023},
  publisher={IOP Publishing}
}

@article{BB_paper,
  title={Studies in astronomical time series analysis. VI. Bayesian Block representations},
  author={Scargle, Jeffrey D and Norris, Jay P and Jackson, Brad and Chiang, James},
  journal={The Astrophysical Journal},
  volume={764},
  number={2},
  pages={167},
  year={2013},
  publisher={IOP Publishing}
}

@article{Crab_2024,
  title={Performance of the HAWC observatory and TeV gamma-ray measurements of the Crab Nebula with improved extensive air shower reconstruction algorithms},
  author={Albert, A and Alfaro, R and Alvarez, C and Andr{\'e}s, A and Arteaga-Vel{\'a}zquez, JC and Rojas, D Avila and Solares, HA Ayala and Babu, R and Belmont-Moreno, E and Bernal, A and others},
  journal={The Astrophysical Journal},
  volume={972},
  number={2},
  pages={144},
  year={2024},
  publisher={IOP Publishing}
}

@article{hawcML,
  title={HAWC Performance Enhanced by Machine Learning in Gamma-Hadron Separation},
  author={Alfaro, R and Alvarez, C and Andr{\'e}s, A and Anita-Rangel, E and Araya, M and Arteaga-Vel{\'a}zquez, JC and Rojas, D Avila and Solares, HA and Babu, R and Bangale, P and others},
  journal={arXiv preprint arXiv:2506.18277},
  year={2025}
}

@article{amon,
  title={AMON Multimessenger Alerts: Past and Future},
  author={Solares, Hugo Alberto Ayala},
  journal={Galaxies},
  volume={7},
  number={1},
  pages={19},
  year={2019},
  publisher={MDPI AG}
}

@article{wilks1938large,
  title={The large-sample distribution of the likelihood ratio for testing composite hypotheses},
  author={Wilks, Samuel S},
  journal={The annals of mathematical statistics},
  volume={9},
  number={1},
  pages={60--62},
  year={1938},
  publisher={JSTOR}
}

@article{3ML,
  title={The multi-mission maximum likelihood framework (3ML)},
  author={Vianello, Giacomo and others},
  journal={arXiv preprint arXiv:1507.08343},
  year={2015}
}

@inproceedings{hal,
  title={Characterizing gamma-ray sources with HAL (HAWC Accelerated likelihood) and 3ML},
  author={Abeysekara and others},
  booktitle={37th International Cosmic Ray Conference.},
  year={2022}
}

@article{franceschini2008extragalactic,
  title={Extragalactic optical-infrared background radiation, its time evolution and the cosmic photon-photon opacity},
  author={Franceschini, Alberto and Rodighiero, Giulia and Vaccari, Mattia},
  journal={Astronomy \& Astrophysics},
  volume={487},
  number={3},
  pages={837--852},
  year={2008},
  publisher={EDP Sciences}
}

@article{BB_code,
  title={Statistical properties of flux variations in blazar light curves at GeV and TeV energies},
  author={Wagner, Sarah M and Burd, Paul R and Dorner, Daniela and Mannheim, Karl and Buson, Sara and Gokus, Andrea and Madejski, Greg and Scargle, Jeffrey D and Arbet-Engels, Axel and Baack, Dominik and others},
  journal={arXiv preprint arXiv:2110.14797},
  year={2021}
}

@article{hawc_mrk_steadystate,
  title={Long-term spectra of the blazars Mrk 421 and Mrk 501 at TeV energies seen by HAWC},
  author={Albert, A and Alfaro, R and Alvarez, C and Camacho, JR Angeles and Arteaga-Vel{\'a}zquez, JC and Arunbabu, KP and Rojas, D Avila and Solares, HA Ayala and Baghmanyan, V and Belmont-Moreno, E and others},
  journal={The Astrophysical Journal},
  volume={929},
  number={2},
  pages={125},
  year={2022},
  publisher={IOP Publishing}
}

@article{hawc-nim-paper,
title = {The High-Altitude Water Cherenkov (HAWC) observatory in México: The primary detector},
journal = {Nuclear Instruments and Methods in Physics Research Section A: Accelerators, Spectrometers, Detectors and Associated Equipment},
volume = {1052},
pages = {168253},
year = {2023},
issn = {0168-9002},
doi = {https://doi.org/10.1016/j.nima.2023.168253},
url = {https://www.sciencedirect.com/science/article/pii/S0168900223002437},
author = {A.U. Abeysekara and A. Albert and R. Alfaro and C. Alvarez and J.D. Álvarez and M. Araya and J.C. Arteaga-Velázquez and K.P. Arunbabu and D. Avila Rojas and H.A. Ayala Solares and R. Babu and A.S. Barber and A. Becerril and E. Belmont-Moreno and S.Y. BenZvi and O. Blanco and J. Braun and C. Brisbois and K.S. Caballero-Mora and J.I. Cabrera Martínez and T. Capistrán and A. Carramiñana and S. Casanova and M. Castillo and O. Chaparro-Amaro and U. Cotti and J. Cotzomi and S. Coutiño {de León} and E. {de la Fuente} and C. {de León} and T. {De Young} and R. Diaz Hernandez and B.L. Dingus and M.A. DuVernois and M. Durocher and J.C. Díaz-Vélez and R.W. Ellsworth and K. Engel and C. Espinoza and K.L. Fan and K. Fang and B. Fick and H. Fleischhack and J.L. Flores and N. Fraija and J.A. García-González and G. Garcia-Torales and F. Garfias and G. Giacinti and H. Goksu and M.M. González and A. González-Muñoz and J.A. Goodman and J.P. Harding and E. Hernandez and S. Hernandez and J. Hinton and B. Hona and D. Huang and F. Hueyotl-Zahuantitla and C.M. Hui and T.B. Humensky and P. Hüntemeyer and A. Iriarte and A. Imran and A. Jardin-Blicq and V. Joshi and S. Kaufmann and D. Kieda and G.J. Kunde and A. Lara and R. Lauer and W.H. Lee and D. Lennarz and H. León Vargas and J.T. Linnemann and A.L. Longinotti and G. Luis-Raya and J. Lundeen and K. Malone and V. Marandon and A. Marinelli and O. Martinez and I. Martínez-Castellanos and J. Martínez-Castro and H. Martínez-Huerta and J.A. Matthews and P. Miranda-Romagnoli and T. Montaruli and J.A. Morales-Soto and E. Moreno and M. Mostafá and A. Nayerhoda and L. Nellen and M. Newbold and M.U. Nisa and R. Noriega-Papaqui and T. Oceguera-Becerra and L. Olivera-Nieto and N. Omodei and A. Peisker and Y. Pérez Araujo and E.G. Pérez-Pérez and E. Ponce and J. Pretz and C.D. Rho and D. Rosa-González and E. Ruiz-Velasco and H. Salazar and D. Salazar-Gallegos and F. Salesa Greus and A. Sandoval and M. Schneider and H. Schoorlemmer and J. Serna-Franco and G. Sinnis and A.J. Smith and Y. Son and K. Sparks Woodle and R.W. Springer and I. Taboada and A. Tepe and O. Tibolla and K. Tollefson and I. Torres and R. Torres-Escobedo and R. Turner and F. Ureña-Mena and T.N. Ukwatta and E. Varela and M. Vargas-Magaña and L. Villaseñor and X. Wang and I.J. Watson and F. Werner and S. Westerhoff and E. Willox and I. Wisher and J. Wood and G.B. Yodh and D. Zaborov and A. Zepeda and H. Zhou},

}
\bibliographystyle{aasjournalv7}

\end{document}